\documentclass[pra,nobalancelastpage,twocolumn,superscriptaddress,showpacs,nofootinbib]{revtex4-1}

\usepackage{xcolor,amsthm,amsmath,amsfonts,graphicx,bm,amssymb,subfigure}
\usepackage{epstopdf}
\usepackage{ulem}
\usepackage{soul}

\newcommand{\ket}[1]{|#1\rangle}
\newcommand{\bra}[1]{\langle #1|}

\newcommand{\Ham}{\mathcal H}

\newcommand{\MyTitle}{Dissipative topological superconductors in number-conserving systems}

\date{\today}
\begin{document}

\title{\MyTitle}

\author{Fernando Iemini}
\affiliation{NEST, Scuola Normale Superiore \& Istituto Nanoscienze-CNR, I-56126 Pisa, Italy}
\affiliation{Departamento de F\'isica - ICEx - Universidade Federal de Minas Gerais, Belo Horizonte - MG, Brazil}
\affiliation{Kavli Institute for Theoretical Physics, University of California, Santa Barbara, CA 93106-4030, USA}

\author{Davide Rossini}
\affiliation{NEST, Scuola Normale Superiore \& Istituto Nanoscienze-CNR, I-56126 Pisa, Italy}
\affiliation{Kavli Institute for Theoretical Physics, University of California, Santa Barbara, CA 93106-4030, USA}

\author{Rosario Fazio}
\affiliation{ICTP, Strada Costiera 11, I-34151 Trieste, Italy}
\affiliation{NEST, Scuola Normale Superiore \& Istituto Nanoscienze-CNR, I-56126 Pisa, Italy}
\affiliation{Kavli Institute for Theoretical Physics, University of California, Santa Barbara, CA 93106-4030, USA}

\author{Sebastian Diehl}
\affiliation{Institute of Theoretical Physics, University of Cologne, D-50937 Cologne, Germany}
\affiliation{Institute of Theoretical Physics, TU Dresden, D-01062 Dresden, Germany}
\affiliation{Kavli Institute for Theoretical Physics, University of California, Santa Barbara, CA 93106-4030, USA}

\author{Leonardo Mazza}
\affiliation{D\'epartement de Physique, Ecole Normale Sup\'erieure / PSL Research University, CNRS, 24 rue Lhomond, F-75005 Paris, France }
\affiliation{NEST, Scuola Normale Superiore \& Istituto Nanoscienze-CNR, I-56126 Pisa, Italy}
\affiliation{Kavli Institute for Theoretical Physics, University of California, Santa Barbara, CA 93106-4030, USA}

\date{\today}

\begin{abstract}
  We discuss the dissipative preparation of p-wave superconductors in number-conserving one-dimensional fermionic systems. We focus on two setups: the first one entails a single wire coupled to a bath, whereas in the second one the environment is connected to a two-leg ladder.
  Both settings lead to stationary states which feature the bulk properties of a p-wave superconductor, identified in this number-conserving setting through the long-distance behavior of the proper p-wave correlations. 
  The two schemes differ in the fact that the steady state of the single wire is not characterized by topological order, whereas the two-leg ladder hosts Majorana zero modes, which are decoupled from damping and exponentially localized at the edges.
  Our analytical results are complemented by an extensive numerical study of the steady-state properties, of the asymptotic decay rate and of the robustness of the protocols.
\end{abstract}

\maketitle


\section{Introduction}

Topological quantum computation has recently emerged as one of the most intriguing paradigms 
for the storage and manipulation of quantum information~\cite{Nayak_2008, Pachos_2012}.
The defining features of topological order, namely the existence of degenerate ground states which 
(i) share the same thermodynamic properties and (ii) can only be distinguished by a global measurement, 
portend for a true many-body protection of quantum information. 
Additionally, the non-Abelian anyons which typically appear in these models are crucial 
for the active manipulation of the information, to be accomplished through 
their adiabatic braiding~\cite{Kitaev_2003, Dennis_2002}.

Among the several systems featuring topological order, 
free p-wave superconducting systems 
with symmetry protected topological properties
have lately attracted a significant amount of attention~\cite{Alicea_2012, Beenakker_2013, DasSarma_2015}.
On the one hand, they are exactly-solvable fermionic models which help building
a clear physical intuition of some aspects of topological order~\cite{Kitaev_2001, Kitaev_2006}.
On the other one, they are physically relevant, and several articles have recently reported 
experimental evidences to be linked to p-wave-like superconductors featuring zero-energy 
Majorana modes~\cite{Mourik_2012, Das_2012, Churchill_2013, Finck_2013, NadjPerge_2014}. 

Whereas up to now these experimental results have been obtained in solid-state setups, 
it is natural to ask whether such physics might as well be observed in cold 
atomic gases~\cite{Radzihovsky_2007}, which owing to their well-controlled microscopic physics should allow 
for a more thorough understanding of these peculiar phases of matter.
Important theoretical efforts have thus proposed a variety of schemes which exploit 
in different ways several properties of such 
setups~\cite{Sato_2009, DasSarma_2011, Jiang_2011, Diehl_2011, Nascimbene_2013, Kraus_2013, Buhler_2014}.

Among these ideas, that of a dissipative preparation of interesting many-body quantum states~\cite{Diehl_2008, Verstraete_2009} 
is particularly appealing: rather than suffering from some unavoidable open-system dynamics,
such as three-body losses or spontaneous emission, one tries to take advantage of it 
(see Refs.~\cite{Roncaglia_2010, Diehl_2011, Bardyn_2013, Budich_2015,Weimer_2010}
for the case 
of states with topological order, such as p-wave superconductors). 
The key point is the engineering of an environment that in the long-time limit drives 
the system into the desired quantum state. This approach has the remarkable advantage of being 
a workaround to the ultra-low temperatures necessary for the observation of important 
quantum phenomena which constitute a particularly severe obstacle in fermionic systems. 
The trust is thus that the mentioned ``non-equilibrium cooling'' may open the path towards the experimental investigation of currently unattainable states, e.g. characterized by p-wave superconductivity. 

In this article we discuss the dissipative engineering of a p-wave superconductor 
with a fixed number of particles, an important constraint in cold-atom experiments.
We consider two different setups:
(i) A single quantum wire,  introduced in Ref.~\cite{Diehl_2011}; 
this system displays the typical features of a p-wave superconductor but it is not topological in its number conserving variant.
(ii) A two-leg ladder~\cite{Fidkowski_2011, Sau_2011, Cheng_2011, Kraus_2013, Ortiz_2014, Iemini_2015, Lang_2015}, supporting a dissipative dynamics which entails a two-dimensional steady state space characterized by p-wave superconducting order with boundary Majorana modes for every fixed particle number.

We identify the p-wave superconducting nature of the steady states by studying the proper correlators, which saturate to a finite value in the long distance limit.
Their topological properties are best discussed using a mathematical connection between dark states of the Markovian dynamics and ground states of a suitable parent Hamiltonian.
In both setups we demonstrate that the dissipative gap closes at least polynomially in the system size and thus that the typical decay time to the steady state diverges in the thermodynamic limit. 
This contrasts with the case where number conservation is not enforced. In this case typically the decay time is finite in the thermodynamic limit~\cite{Diehl_2011, Bardyn_2013}, and reflects the presence of dynamical slow modes related to the particle-number conservation~\cite{Cai_2013,Hohenberg_1977}, which also exist in non-equilibrium systems (see also Ref.~\cite{Kastoryano_2012,Buchhold_2015,Caspar_2015}).

Our exact analytical findings are complemented by a numerical study based on a matrix-product-operator 
representation of the density matrix~\cite{Verstraete_2004, Zwolak_2004}, 
one of the techniques for open quantum systems which are recently attracting an increasing 
attention~\cite{Orus_2008, Prosen_2009, Hartmann_2009, Daley_2014, Cui_2015, Biella_2015, Finazzi_2015, Mascarenhas_2015, Werner_2015}.
These methods are employed to test the robustness of these setups to perturbations, which is thoroughly discussed.

The article is organized as follows: in Sec.~\ref{sec:I} we review the key facts behind the idea 
of dissipative state preparation using the dark states of a many body problem, and exemplify them 
recalling the problem studied in Ref.~\cite{Diehl_2011}. A simple criterion for signalling the divergence of the decay-time with the system size is also introduced. 
In Sec.~\ref{sec:N:1} we present the exact analytical study of the single-wire protocol, 
and in Sec.~\ref{sec:Numerics1} a numerical analysis complements the previous discussion with the characterization of the robustness to perturbations of these setups. 
In Sec.~\ref{sec:N:2} we discuss the protocol based on the ladder geometry. 
Finally, in Sec.~\ref{sec:conc} we present our conclusions.

\section{Dissipative state preparation of Majorana fermions: known facts}
\label{sec:I}

\subsection{Dark states and parent Hamiltonian of Markovian dynamics} \label{sec:I:A}

The dissipative dynamics considered in this article is Markovian and, in the absence
of a coherent part, can be cast in the following Lindblad form:
\begin{equation}
  \frac{\partial}{\partial t}\hat \rho =
  \mathcal L[\hat \rho] =
  \sum_{j=1}^{m} \left[ \hat L_j \hat \rho \hat L_j^\dagger - \frac 12 \{ \hat L_j^\dagger \hat L_j, \hat \rho \} \right],
  \label{eq:master:equation}
\end{equation}
where $\mathcal L$ is the so-called Lindbladian super-operator and the $\hat L_j$ are the (local) 
Lindblad operators.
We now discuss a fact which will be extensively used in the following.
Let us assume that a pure state $\ket{\Psi}$ exists, with the property:
\begin{equation}
  \hat L_j \ket{\Psi} = 0; \quad \forall \, j = 1,\ldots,m.
  \label{eq:dark:state}
\end{equation}
A simple inspection of Eq.~\eqref{eq:master:equation} shows that $\ket{\Psi}$ 
is a steady state of the dynamics, and it is usually referred to as \textit{dark state}.
Although the existence of a state satisfying Eq.~\eqref{eq:dark:state} is usually not guaranteed, 
in this article we will mainly consider master equations which enjoy this property.

A remarkable feature of dark states is that they can be searched through the minimization 
of a \textit{parent Hamiltonian}. Let us first observe that Eq.~\eqref{eq:dark:state} implies that 
$\bra {\Psi} \hat L_j^\dagger \hat L_j \ket{\Psi} = 0$ and since every operator 
$\hat L_j^\dagger \hat L_j$ is positive semi-definite, $\ket {\Psi}$ minimizes it.
Consequently, $\ket{\Psi}$ is a ground state of the parent Hamiltonian: 
\begin{equation}
  \hat {\mathcal H}_{p} = \sum_{j=1}^m \hat L^\dagger_j \hat L_j.
  \label{eq:parent:Hamiltonian}
\end{equation}
Conversely, every zero-energy ground state $\ket{\Phi}$ of Hamiltonian~\eqref{eq:parent:Hamiltonian} 
is a steady state of the dynamics~\eqref{eq:master:equation}.
Indeed, $\hat {\mathcal H}_p \ket {\Phi} = 0$ implies that 
$\bra {\Phi} \hat L_j^\dagger \hat L_j \ket{\Phi} = 0$ for all $j =1,\ldots,m$.
The last relation means that the norm of the states $\hat L_j \ket{\Phi}$ is zero, 
and thus that the states themselves are zero: $\hat L_j \ket{\Phi}=0$. 
As we have already shown, this is sufficient to imply that $\ket {\Phi}$ is a steady-state of the dynamics.

In order to quantify the typical time-scale of the convergence to the steady state, 
it is customary to consider the right eigenvalues of the super-operator $\mathcal L$, 
which are defined through the secular equation $\mathcal L [\hat \theta_\lambda] = \lambda \hat \theta_{\lambda}$.
The asymptotic decay rate for a finite system is defined as 
\begin{equation}
  \lambda_{\rm ADR} = \inf_{\substack{\lambda \text{ is eigenvalue of } \mathcal L\\ \Re(\lambda) \neq 0}} \{ -\Re (\lambda)\}.
  \label{eq:ADR}
\end{equation}
The minus sign in the previous equation follows from the fact that the real part of the eigenvalues 
of a Lindbladian super-operator satisfy the following inequality: $\Re(\lambda) \leq 0$. 

Remarkably, for every eigenvalue $\xi$ of $\hat {\mathcal H}_p$ there is an eigenvalue 
$\lambda = - \xi/2$ of $\mathcal L$ which is at least two-fold degenerate. 
Indeed, given the state $\ket {\psi_\xi}$ such that $\hat {\mathcal H}_p \ket{\psi_\xi} = \xi \ket{\psi_\xi}$, 
the operators made up of the dark state $\ket{\Psi}$ and of $\ket{\psi_\xi}$
\begin{equation}
  \hat \theta_{- \xi/2}^{(1)} = \ket{\Psi} \hspace{-0.05cm} \bra{\psi_\xi},
  \quad
  \hat \theta_{- \xi/2}^{(2)} = \ket{\psi_\xi} \hspace{-0.05cm} \bra{\Psi}
\end{equation}
satisfy the appropriate secular equation.
This has an important consequence: if $\hat {\mathcal H}_p$ is gapless, 
then $\lambda_{\rm ADR} \xrightarrow{L \to \infty} 0$ in the thermodynamic limit, where $L$ is the
size of the system. Indeed: 
\begin{equation}
  0 < \lambda_{\rm ADR} \leq \frac \xi2,
\end{equation}
for every eigenvalue $\xi$ of $\hat {\mathcal H}_p$; if $\xi$ closes as $L^{-\alpha}$ ($\alpha>0$), 
then the dissipative gap closes at least polynomially in the system size.
Note that this argument also implies that if $\mathcal L$ is gapped, 
then the parent Hamiltonian is gapped as well.

It is important to stress that the spectral properties of the parent Hamiltonian $\hat {\mathcal H}_p$
do not contain all the information concerning the long-time dissipative dynamics.
As an example, let us assume that the Markovian dynamics in Eq.~\eqref{eq:master:equation} 
(i) supports at least one dark state and (ii) has an associated parent Hamiltonian which is gapped.
If the Lindblad operators are Hermitian, then the fully-mixed state is a steady state 
of the master equation too. The presence of such stationary state is not signaled 
by the parent Hamiltonian, which is gapped and only detects the pure steady states of the dynamics.

Whereas the some of the above relations have been often pointed out in the 
literature~\cite{Diehl_2008, Verstraete_2009}, to the best of our knowledge the remarks 
on the relation between the spectral properties of $\mathcal L$ and $\hat {\mathcal H}_p$ are original.

\subsection{The Kitaev chain and the dissipative preparation of its ground states} \label{sec:I:B}

Let us now briefly review the results in Ref.~\cite{Diehl_2011} and use them to exemplify 
how property~\eqref{eq:dark:state} can be used as a guideline for dissipative state preparation in the number non-conserving case. This will be valuable for our detailed studies of its number conserving variant below.

The simplest model displaying zero-energy unpaired Majorana modes is the one-dimensional 
Kitaev model at the so-called ``sweet point''~\cite{Kitaev_2001}:
\begin{equation}
  \hat {\mathcal H}_{\rm K} = 
  -J \sum_j \left[ \hat a^\dagger_j \hat a_{j+1} + \hat a_j \hat a_{j+1} + \mathrm{H.c.} \right],
  \quad J>0,
\end{equation}
where the fermionic operators $\hat a_j^{(\dagger)}$ satisfy canonical anticommutation relations 
and describe the annihilation (creation) of a spinless fermion at site $j$.
The model can be solved with the Bogoliubov-de-Gennes transformation, and, when considered 
on a chain of length $L$ with open boundaries, it takes the form:
\begin{equation}
  \hat {\mathcal H}_{\rm K} = E_0
  + \frac J2 \sum_{j=1}^{L-1} \hat {\ell}_j^\dagger \hat \ell_j \, ,
  \label{eq:Kitaev:Hamiltonian}
\end{equation}
with
\begin{eqnarray}
\hat \ell_j &=& \hat C_j^\dagger + \hat A_j, \\ 
\hat C^\dag_j &=& \hat a^\dag_j+\hat a^\dag_{j+1}, \quad \hat A_j = \hat a_j - \hat a_{j+1}.
\end{eqnarray}

The ground state has energy $E_0$ and is two-fold degenerate: there are two linearly 
independent states $\ket {\psi_e}$ and $\ket {\psi_o}$ which satisfy: 
\begin{equation}
  \hat \ell_j \ket{\psi_\sigma}=0; \quad 
  \forall \, j = 1,\ldots, L-1;
  \qquad \sigma =e,o.
  \label{eq:key:property}
\end{equation}
The quantum number distinguishing the two states is the parity of the number of fermions, 
$\hat P = (-1)^{\sum \hat a_j^\dagger \hat a_j}$, which is a symmetry of the model 
(the subscripts $e$ and $o$ stand for \textit{even} and \textit{odd}).
Both states $\ket{\psi_{\sigma}}$ are p-wave superconductors, as it can be explicitly proven 
by computing the expectation value of the corresponding order parameter:
\begin{equation}
  \bra{\psi_\sigma} \hat a_j \hat a_{j+1} \ket{\psi_\sigma} 
  \xrightarrow{L \rightarrow \infty} \frac{1}{4}.
  \label{eq:order:par}
\end{equation}

It is thus relevant to develop a master equation which features $\ket{\psi_e}$ and $\ket {\psi_o}$ 
as steady states of the dynamics~\cite{Diehl_2011, Bardyn_2013}.
Property~\eqref{eq:key:property} provides the catch: upon identification of the $\hat \ell_j$ operators 
with the Lindblad operators of a Markovian dynamics, Eq.~\eqref{eq:dark:state} ensures that 
the states $\ket{\psi_\sigma}$ are steady states of the dynamics and that in the long-time limit 
the system evolves into a subspace described in terms of p-wave superconducting states.
This becomes particularly clear once it is noticed that the parent Hamiltonian of this Markov process 
coincides with $\hat {\mathcal H}_{\rm K}$ in Eq.~\eqref{eq:Kitaev:Hamiltonian} apart from an additive constant.

Let us conclude mentioning that the obtained dynamics satisfies an important physical requirement, 
namely \textit{locality}. 
The Lindblad operators $\hat \ell_j$ only act on two neighboring fermionic modes; 
this fact makes the dynamics both physical and experimentally feasible.
On the other hand, they do not conserve the number of particles, 
thus making their engineering quite challenging with cold-atom experiments. 
The goal of this article is to provide dissipative schemes with Lindblad operators 
which commute with the number operator and feature the typical properties of a p-wave superconductor.

\section{Single wire: Analytical results}
\label{sec:N:1}

The simplest way to generalize the previous results to systems where the number of particles is conserved 
is to consider the master equation induced by the Lindblad operators~\cite{Diehl_2011, Bardyn_2013}:
\begin{equation}
  \hat L'_j = \hat C_j^\dagger \hat A_j,
  \quad \forall \, j =1,\ldots,L-1,
  \label{eq:Lind:N:1}
\end{equation}
for a chain with hard-wall boundaries and spinless fermions:
\begin{equation}
  \frac{\partial}{\partial t}\hat \rho =
  \mathcal L'[\hat \rho] = \gamma
  \sum_{j=1}^{L-1} \left[ \hat L_j' \hat \rho \hat L_j'^\dagger - \frac 12 \{ \hat L_j'^\dagger \hat L'_j, \hat \rho \} \right];
  \quad \gamma > 0;
  \label{eq:master:equation:N:1}
\end{equation}
 where $\gamma$ is the damping rate.
This Markovian dynamics has already been considered in Refs.~\cite{Diehl_2011, Bardyn_2013}. 
Using the results presented in Ref.~\cite{Iemini_2015}, where the parent Hamiltonian related to the dynamics in Eq.~\eqref{eq:master:equation:N:1} is considered, it is possible to conclude that for a chain with periodic boundary conditions (i) there is a unique dark state for every particle number density $\nu = N/L$, and (ii) this state is a p-wave superconductor.
A remarkable point is that the $\hat L'_j$ are local and do not change the number of particles: 
their experimental engineering is discussed in Ref.~\cite{Diehl_2011}, see also \cite{MullerReview}.

Here we clarify that for the master equation for a single wire with hard-wall boundaries, the steady state is not topological and does not feature Majorana edge physics, although they still display the bulk properties of a p-wave superconductor (instead, the two-wire version studied below \textit{has} topological properties associated to dissipative Majorana zero modes).
The asymptotic decay rate of the master equation is also characterized. 
An extensive numerical study of the stability of this protocol is postponed to Sec.~\ref{sec:Numerics1}.

\subsection{Steady states}

In order to characterize the stationary states of the dynamics, 
let us first observe that Eq.~\eqref{eq:key:property} implies~\cite{Iemini_2015}
\begin{equation}
  \hat C^\dagger_j \ket{\psi_\sigma} = - \hat A_j \ket{\psi_\sigma}, 
\end{equation}
so that:
\begin{equation}
  \hat L'_j \ket {\psi_\sigma} = 
  \hat C_j^\dagger \hat A_j \ket{\psi_\sigma} =
  - \hat C_j^\dagger \hat C_j^\dagger \ket{\psi_\sigma} = 0.
\end{equation}
Thus, $\ket{\psi_\sigma}$ are steady states of the dynamics.
Let us define the states 
\begin{equation}
  \ket{\psi_N} = \hat \Pi_N \ket{\psi_\sigma}, 
  \label{eq:proj:N:stst}
\end{equation}
where $\hat \Pi_N$ is the projector onto the subspace of the global Hilbert (Fock) space 
with $N$ fermions ($\hat \Pi_N \ket{\psi_\sigma} = 0$ when the parity of $N$ differs from $\sigma$ 
and thus we avoid the redundant notation $\ket{\psi_{\sigma,N}}$). 
Since $[\hat L'_j, \hat N]=0$, where $\hat N = \sum_j\hat a^\dagger_j \hat a_j$ is the particle-number 
operator, it holds that $\hat L'_j \ket{\psi_N}=0$ for all $j = 1, \ldots, L-1$ 
and thus the $\ket{\psi_N}$ are dark states.
Let us show that there is only one dark state $\ket{\psi_N}$ once the value of $N$ is fixed.
To this end, we consider the parent Hamiltonian~\eqref{eq:parent:Hamiltonian} associated 
to the Lindblad operators~\eqref{eq:Lind:N:1}:
\begin{equation}
  \hat {\mathcal H}_{p}' \! = \!
   2 J \sum_{j=1}^{L-1} \! \left[ 
    \hat n_j \! + \! \hat n_{j+1} \! - \! 2 \hat n_j \hat n_{j+1} 
    \! - \! \hat a_{j+1}^\dagger \hat a_{j} \! - \! \hat a_j^\dagger \hat a_{j+1}
    \right] \! , \; 
\end{equation}
where $\hat n_j \equiv \hat a_j^\dagger \hat a_j$ 
and $J>0$ is a typical energy scale setting the units of measurement.
Upon application of the Jordan-Wigner transformation, the model $\hat {\mathcal H}_p'$ 
is unitarily equivalent to the following spin-$1/2$ chain model:
\begin{equation}
  \hat {\mathcal H}_{p,\rm spin}' = J
  \sum_{j=1}^{L-1} \left[ 1 
    + \hat \sigma_j^x \hat \sigma_{j+1}^x
    + \hat \sigma_j^y \hat \sigma_{j+1}^y
    - \hat \sigma_j^z \hat \sigma_{j+1}^z \right] ,
\end{equation}
where $\hat \sigma_j^{\alpha}$ are Pauli matrices.
Apart from a constant proportional to $L-1$, $\hat {\mathcal H}_{p, \rm spin}'$ is the ferromagnetic Heisenberg model.
The particle-number conservation corresponds to the conservation of the total magnetization 
along the $\hat z$ direction. It is a well-known fact that this model has a highly degenerate 
ground state but that there is only one ground state for each magnetization sector, 
both for finite and infinite lattices. 
Thus, this state corresponds to the state $\ket {\psi_N}$ identified above; therefore, 
the possibility that the ground state of $\hat {\mathcal H}_p'$ is two-fold degenerate (as would be required for the existence of Majorana modes) 
for fixed number of fermions and hard-wall boundary conditions is ruled out.

Summarizing, the dynamics induced by the Lindblad operators in~\eqref{eq:Lind:N:1}
conserves the number of particles and drives the system into a quantum state with the properties 
of a p-wave superconductor (in the thermodynamic limit $\ket{\psi_e}$ and $\hat \Pi_N \ket{\psi_e}$ 
have the same bulk properties, as it is explicitly checked in Ref.~\cite{Diehl_2011, Bardyn_2013}, 
but see also the discussion below).
Since the steady states of the system for open boundary conditions are unique, they do not display 
any topological edge property.

\subsection{P-wave superconductivity}

Let us explicitly check that the states $\ket{\psi_N}$ have the properties of a p-wave superconductor. 
Since each state has a definite number of fermions, the order parameter defined in Eq.~\eqref{eq:order:par} 
is zero by symmetry arguments. 
In a number-conserving setting, we thus rely on the p-wave pairing correlations:
\begin{equation}
  G^{(p)}_{j,l} = \bra{\psi_N} \hat O_j^{(p)\dagger} \hat O^{(p)}_l \ket{\psi_N} 
  = \bra{\psi_N} \hat a_j^\dagger \hat a_{j+1}^\dagger \hat a_{l+1} \hat a_{l} \ket{\psi_N} .
  \label{eq:observable:pairing}
\end{equation}
If in the long-distance limit, $|l-j| \to \infty$, the expectation value saturates to a finite value 
or shows a power-law behavior, the system displays p-wave superconducting (quasi-)long-range order. 
If the decay is faster, e.g. exponential, the system is disordered.

In this specific case, the explicit calculation shows a saturation at large distance 
(see also Ref.~\cite{Iemini_2015}):
\begin{equation}
  G^{(p)}_{j,l} \xrightarrow{|j-l| \to \infty} \nu^2 (1-\nu)^2
  \label{eq:order:corr}
\end{equation}
in the thermodynamic limit.
The saturation to a finite value captures the p-wave superconducting nature of the states.
Note that the breaking of a continuous symmetry in a one-dimensional system signaled 
by Eq.~\eqref{eq:order:corr} is a non-generic feature: a perturbation of Hamiltonian $\hat {\mathcal H}_p'$ 
would turn that relation into a power-law decay to zero as a function of $|j-l|$ (see Ref.~\cite{Iemini_2015} for an explicit example).

\subsection{Dissipative gap}

An interesting feature of $\hat {\mathcal H}_{p,\rm spin}'$ is that it is gapless; the gap closes 
as $L^{-2}$ due to the fact that the low-energy excitations have energy-momentum relation $\omega_q\sim q^2$, 
as follows from well-known properties of the ferromagnetic Heisenberg model. 
The Jordan-Wigner transformation conserves the spectral properties and thus $\hat {\mathcal H}_{p}'$ 
is also gapless. Thus, according to the discussion in Sec.~\ref{sec:I:A}, 
the asymptotic decay rate $\lambda_{\rm ADR}'$ associated to the Lindbladian $\mathcal L'$ 
closes in the thermodynamic limit. This is true both for periodic and hard-wall boundary conditions.

This fact has two important consequences.
The first is that the dissipative preparation of a fixed-number p-wave superconductor through 
this method requires at least a typical time $\tau'$ that scales like $L^2$. 
In Sec.~\ref{sec:Numerics1} we numerically confirm this polynomial scaling. 
Although this requires an effort which is polynomial in the system size, and which is thus efficient, 
it is a slower dynamical scenario than that of the non-number-conserving dynamics considered 
in Refs.~\cite{Diehl_2011, Bardyn_2013} and summarized in Sec.~\ref{sec:I:B}, 
where $\tau$ does not scale with $L$ (the super-operator $\mathcal L$ in that case is gapped), and thus the approach to stationarity is exponential in time. The difference can be traced to the presence of dynamical slow modes related to exact particle number conservation, a property which is abandoned in the mean field approximation of Refs.~\cite{Diehl_2011, Bardyn_2013}.

The second consequence is that a gapless Lindbladian $\mathcal L$ does not ensure 
an \textit{a priori} stability of the dissipative quantum state preparation. 
Roughly speaking, even a small perturbation $\epsilon \mathcal M'$ ($\epsilon \ll 1$) 
to the Lindbladian $\mathcal L'$ such that the dynamics is ruled by 
$\mathcal L'+ \epsilon \mathcal M'$ has the potential to qualitatively change the physics of the steady-state 
(see Refs.~\cite{Syassen_2008, Li_2014, Ippoliti_2015} for some examples where the presence of a gap 
is exploited for a perturbative analysis of the steady states). This concerns, in particular, the long-distance behavior of correlation functions. 
To further understand this last point, in Sec.~\ref{sec:Numerics1} we have analyzed the effect of several perturbations through numerical simulations.
In the case in which the steady state has topological properties, they may still be robust. We further elaborate on this point in Sec.~\ref{sec:N:2}, where we study the ladder setup.

Notwithstanding the gapless nature of the Lindbladian $\mathcal L'$, we can show that waiting 
for longer times is beneficial to the quantum state preparation.
If we define $p_0(t) = \text{tr} \big[ \hat P_0 \hat \rho(t) \big]$, where $\hat P_0$ is the projector 
onto the ground space of the parent Hamiltonian $\hat {\mathcal H}'_p$, then the following monotonicity property holds:
\begin{equation}
  \frac{\mathrm d}{\mathrm dt} p_0(t) \geq0.
  \label{eq:positive:derivative}
\end{equation}
Indeed, $\frac{\mathrm d}{\mathrm dt} p_0(t) = \text{tr} \big[ \hat P_0 \mathcal L'[\hat \rho(t)] \big] =
\text{tr} \big[ \mathcal L'^*[\hat P_0] \hat \rho(t) \big]$,
where $\mathcal L'^*$ is the adjoint Lindbladian.
It is easy to see that $\mathcal L'^*[\hat P_0] =  \gamma \sum_j \hat L_j'^{\dagger} \hat P_0 \hat L'_j$, 
which is a non-negative operator because for any state $\ket{\phi}$ it holds that:
\begin{align}
  \langle \phi |\mathcal L'^*[\hat P_0]| \phi \rangle 
  =& \gamma \sum_j \langle \phi |\hat L_j'^{\dagger} \hat P_0 \hat L'_j | \phi \rangle = \nonumber \\
=&   \gamma \sum_{j, \alpha} |\langle \psi_\alpha |\hat L'_j | \phi \rangle|^2 >0
\end{align}
where $\{ \ket {\psi_\alpha}\}$ are a basis of the ground space of the parent Hamiltonian $\hat {\mathcal H}'_p$.
If we consider the spectral decomposition of 
$\hat \rho (t) = \sum_\beta p_\beta \ket{\phi_\beta} \hspace{-0.05cm} \bra{\phi_\beta}$, with $p_\beta>0$, 
we obtain Eq.~\eqref{eq:positive:derivative}.

\section{Single wire: Numerical results}
\label{sec:Numerics1}

Although the previous analysis, based on the study of the dark states of the dynamics, has already identified many distinguishing properties of the system, there are several features which lie outside its prediction range. 
Let us list for instance the exact size scaling of the asymptotic decay rate or the resilience of the scheme to perturbations.
In order to complement the analysis of the dissipative dynamics with these data, we now rely on a numerical approach.

The numerical analysis that we are going to present is restricted to systems with hard-wall boundary conditions. 
In order to characterize the time evolution described by the master equation~\eqref{eq:master:equation:N:1}, 
we use two different numerical methods. The first is a Runge-Kutta (RK) integration for systems 
of small size (up to $L=10$)~\cite{bookPress}. 
This method entails an error due to inaccuracies in the numerical integration, 
but the density matrix is represented without any approximation.

On the contrary, the second method, based on a Matrix-Product-Density-Operator (MPDO) representation 
of the density matrix, allows the study of longer systems through an efficient approximation 
of $\hat \rho$~\cite{Verstraete_2004, Zwolak_2004, Prosen_2009}. 
The time evolution is performed through the Time-Evolving Block Decimation (TEBD) algorithm, 
which is essentially based on the Trotter decomposition of the Liouville super-operator $e^{t \mathcal L'}$.
Although this method has been shown to be able to reliably describe problems with 
up to $\sim 100$ sites~\cite{Biella_2015}, in this case we are not able to consider lengths 
beyond $L=22$ because of the highly-entangled structure of the states encountered during the dynamics. 
It is an interesting perspective to investigate whether algorithms based on an MPDO representation 
of the density matrix, which compute the steady state through maximization of the Lindbladian 
super-operator $\mathcal L'$, might prove more fruitful in this context~\cite{Cui_2015, Mascarenhas_2015}.

Finally, we have also performed Exact-Diagonalization (ED) studies of system sizes up to $L=5$ 
in order to access properties of $\mathcal L'$, such as its spectrum, 
which cannot be observed with the time-evolution.

\subsection{Asymptotic decay rate}

\begin{figure}
  \centering
  \includegraphics[scale=0.45]{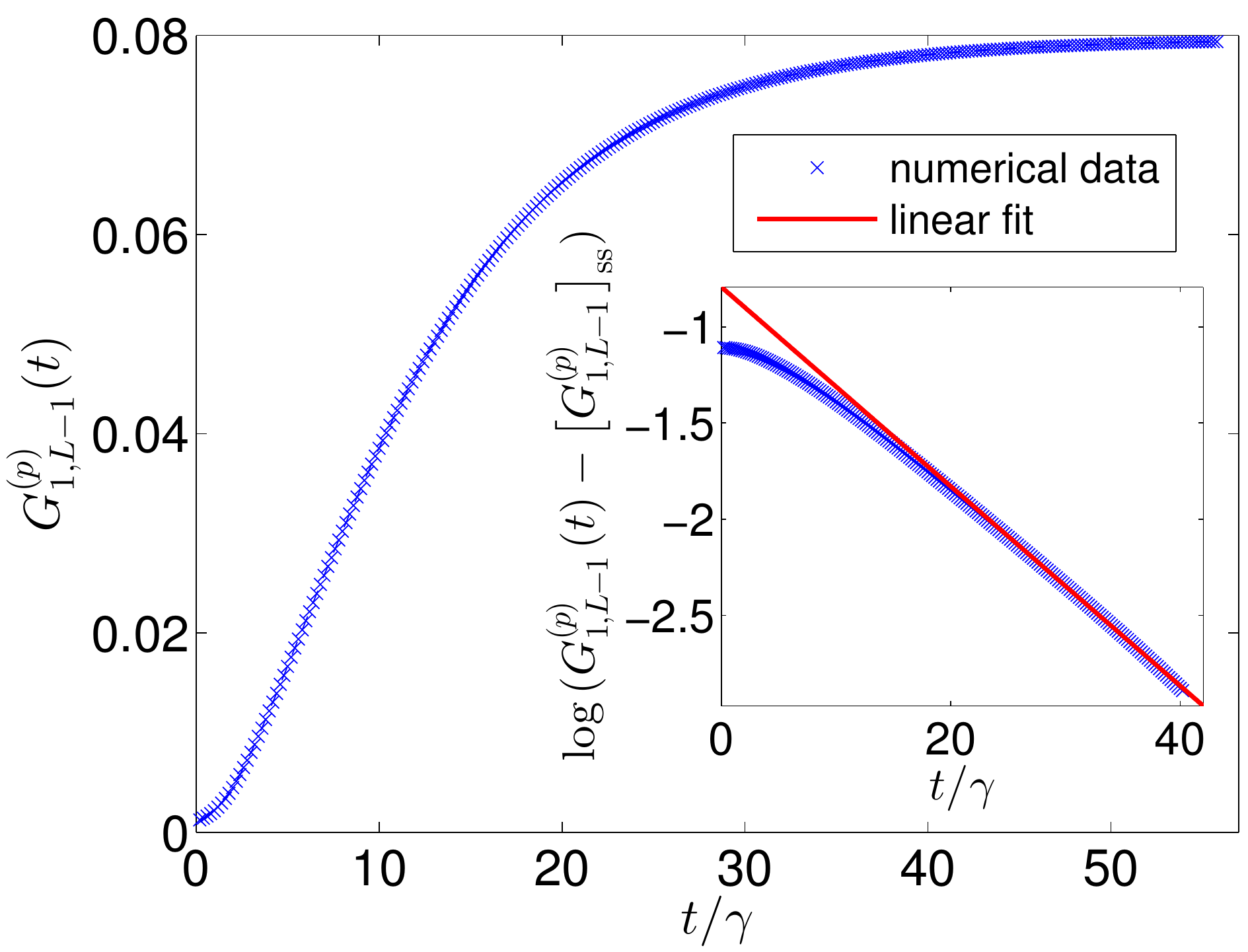}
  \includegraphics[scale=0.45]{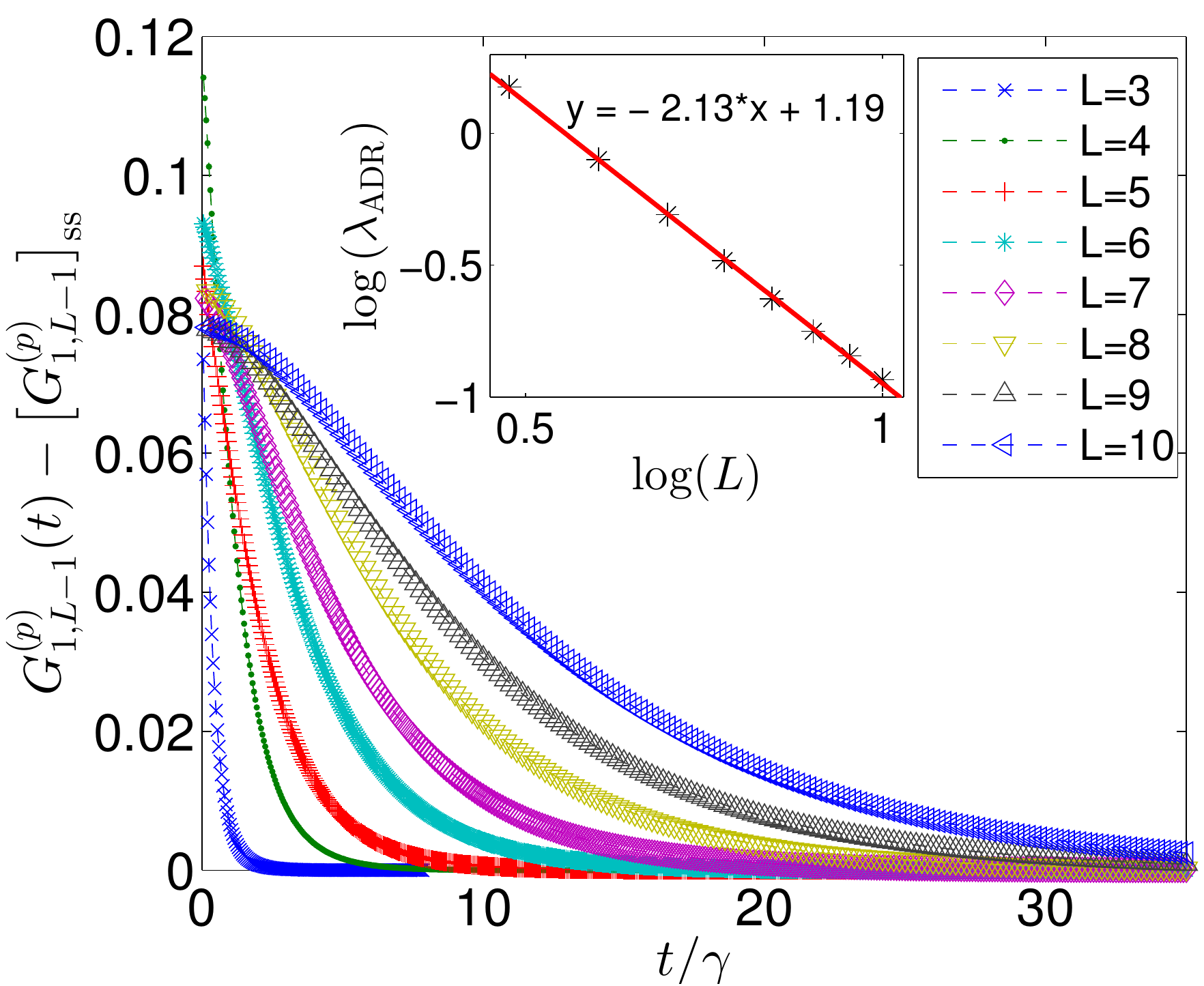}
  \caption{(Color online) (Top) Runge-Kutta time evolution of the pairing correlator $G^{(p)}_{1,L-1}(t)$ 
    for the largest available system size, $L=10$.
    The inset shows that upon subtraction of the steady value, an exponential decay is observed, 
    from which $\lambda_{\rm ADR}$ is extracted.
    (Bottom) Time evolution of $G^{(p)}_{1,L-1}(t) - \big[ G^{(p)}_{1,L-1} \big]_{\rm ss}$ for several system sizes.
    The inset shows the scaling of $\lambda_{\rm ADR}$ with $L$, which is fitted by an algebraic function.
  }
  \label{scaling.diss.gap}
\end{figure} 

Let us first assess that the asymptotic decay rate of the system closes polynomially 
with the system size (from the previous analysis we know that it closes \textit{at least} polynomially.
As we discuss in Appendix~\ref{app:Spectral}, in the asymptotic limit, 
it is possible to represent the expectation value of any observable $\hat A$ as: 
\begin{equation}
  \langle \hat{A}\rangle(t) - \langle \hat{A}\rangle_{\rm ss} \sim \kappa e^{-\lambda_{\rm ADR} t} + \ldots
  \label{eq:steady:observable}
\end{equation}
where $\langle \hat A \rangle(t) = {\rm tr} \big[ \hat A \, \hat \rho(t) \big]$, 
$\langle \hat{A}\rangle_{\rm ss} = \lim_{t \to \infty} \langle \hat{A}\rangle(t)$
and $\kappa$ is a non-universal constant.
The notation $-\lambda_{\rm ADR}$ is due to the fact that $\lambda_{\rm ADR}$ is positive, being defined through the additive inverse of the real part of the eigenvalues, see Eq.~\eqref{eq:ADR}.
It is possible to envision situations where $\kappa=0$ and thus the long-time decay 
is dominated by eigenvalues of $\mathcal L'$ with smaller real part.

The study of the long-time dependence of any observable can be used to extract 
the value of $\lambda_{\rm ADR}$; among all the possible choices, we employ the pairing correlator 
$G^{(p)}_{j,l}(t) = \langle \hat O_j^{(p)\dagger} \hat O^{(p)}_l \rangle(t)$ [see Eq.~\eqref{eq:observable:pairing}]
because of its special physical significance.
In Fig.~\ref{scaling.diss.gap}(top), we consider $L=10$ and plot the time evolution of $G^{(p)}_{j,l}(t)$ 
for $j=1$ and $l = L-1$ (no relevant boundary effects have been observed as far as the estimation 
of $\lambda_{\rm ADR}$ is concerned). The calculation is performed through RK integration of the master equation. 
The initial state of the evolution is given by the ground state of the non-interacting Hamiltonian, 
$\hat {\mathcal H}_0 = - J
\sum_j \hat a^{\dagger}_{j} \hat a_{j+1} + {\rm H.c.}$ 
($N=L/2$ for $L$ even, and $N=(L+1)/2$ for $L$ odd).

In order to benchmark the reliability of the RK integration for getting the steady state,
we compare the expectation value 
of several observables (in particular of pairing correlators) with the exactly-known results 
(Sec.~\ref{sec:N:1} provides the exact wavefunction of the steady state, from which several observables 
can be computed). In all cases the absolute differences are below $10^{-6}$.
Similar results are obtained for smaller system sizes, where it is even possible to compute 
the trace-distance of the RK steady-state from the $\lambda=0$ eigenstate of the Liouvillian computed with ED. 

In the long-time limit, the observable~\eqref{eq:observable:pairing} displays a clear stationary behavior, 
$\big[ G^{(p)}_{j,l} \big]_{\rm ss} = \lim_{\tau \to \infty} G^{(p)}_{j,l} (\tau)$, 
consistently with Eq.~\eqref{eq:steady:observable}. 
Once such stationary value is subtracted, 
it is possible to fit $\lambda_{\rm ADR}$ from the exponential decay of
\begin{equation}
  G^{(p)}_{j,l}(t) - \big[ G^{(p)}_{j,l} \big]_{\rm ss}
  \label{eq:subtraction}
\end{equation}
The subtraction is possible to high precision because the value of $\big[ G^{(p)}_{j,l} \big]_{\rm ss}$ is known 
from the previous analytical considerations.
Moreover, as we have already pointed out, the evolution continues up to times such that 
$G^{(p)}_{j,l} (t)$ differs in absolute terms from 
the analytical value for $\lesssim 10^{-6}$, which makes the whole procedure reliable. 

In Fig.~\ref{scaling.diss.gap}(bottom) we display the quantity in~\eqref{eq:subtraction} 
for various lattice sizes $L$. It is clear that the convergence of the observable requires an amount 
of time which increases with $L$. 
A systematic fit of $\lambda_{\rm ADR}$ for several chain lengths allows for an estimate 
of its dependence on $L$ [see Fig.~\ref{scaling.diss.gap}(bottom)]: the finite-size dissipative gap scales as
\begin{equation}
  \lambda_{\rm ADR}\propto L^{-2.13 \pm 0.05 } \; .
  \label{eq:lambda:ADR:scaling}
\end{equation}
The exact diagonalization (ED) of the Liouvillian up to $L=5$ allows a number of further considerations.
First, the Liouvillian eigenvalues with largest real part ($\Re(\lambda) \lesssim 0$) are independent of the number of particles (the check has been performed for every value of $N=1, ... , 5$).
Second, comparing the ED with the previous analysis, we observe that 
the $\lambda_{\rm ADR}$ in Eq.~\eqref{eq:lambda:ADR:scaling} coincides with the second eigenvalue 
of the Liouvillian, rather than with the first [here the generalized eigenvalues 
are ordered according to the additive inverse of their real part $-\Re(\lambda)$]. 
Numerical inspection of small systems (up to $L=5$) shows that the first excited eigenvalue 
of $\mathcal L'$ is two-fold degenerate and takes the value $- \xi/2$, where $\xi$ is the energy 
of the first excited state of $\hat {\mathcal H}'_p$ (see the discussion in Sec.~\ref{sec:I:A}). 
Our numerics suggests that it does not play any role in this particular dissipative evolution, 
hinting at the fact that the chosen $\hat \rho(0)$ does not overlap with the eigensubspace 
relative to $-\xi/2$. In this case, the value of $\kappa$ in Eq.~\eqref{eq:steady:observable} is zero.

\subsection{Perturbations}

In order to test the robustness of the dissipative scheme for the preparation of a p-wave superconductor, 
we now consider several perturbations of the Lindbladian $\mathcal L'$ of both dissipative and Hamiltonian form. 
The robustness of the dissipative state preparation of the p-wave superconductor is probed 
through the behavior of the correlations $G^{(p)}_{j,l}(t)$, which define such phase.

\subsubsection{Perturbations of the Lindblad operators}

\begin{figure}
  \centering
  \includegraphics[scale=0.45]{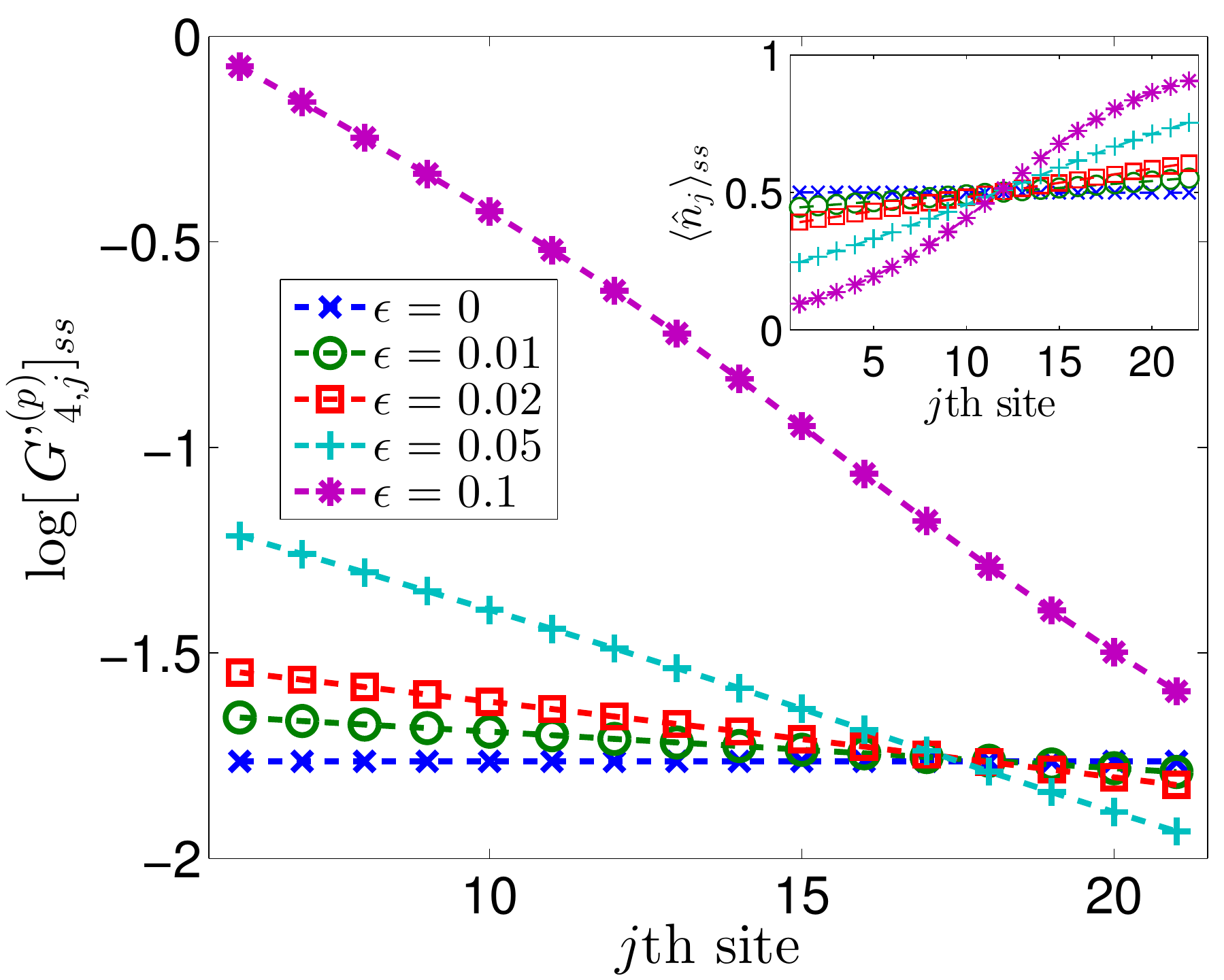} 
  \caption{(Color online) Steady-state values of $\big[ G'^{(p)}_{4,j} \big]_{\rm ss}$  [see Eq.~\eqref{eq:gprime}]
    for a lattice with $L=22$ sites at half-filling, $\nu=1/2$, computed with MPDO for different values 
    of $\epsilon$ in $\hat L'_{j,\epsilon}$ [see Eq.~\eqref{eq:pert:exc}].
    The inset displays the steady-state values of the local number of fermions 
    $\langle \hat n_j \rangle_{\rm ss}$ for the same systems.
}
  \label{excitation.perturbation.L22}
\end{figure}

Let us define the following perturbed Lindblad operator:
\begin{equation}
  \hat L'_{j,\epsilon} = \hat C_{j}^\dagger \hat A_{j,\epsilon};
  \quad 
  \hat A_{j,\epsilon} = \hat a_j-(1-\epsilon)\hat a_{j+1};
  \quad \epsilon \in \mathbb R,
  \label{eq:pert:exc}
\end{equation}
which allows for slight asymmetries in the action of the dissipation 
between sites $j$ and $j+1$. The continuity equation associated to the dynamics, $\partial_t \hat n_i = -(\hat j_i - \hat j_{i-1})$,
is characterized by the following current operator: $\hat j_i =  \hat n_i - (1- \epsilon)^2 \hat n_j + (\epsilon^2 - 2 \epsilon) \hat n_i \hat n_{i+1}$. When $\epsilon \neq 0$, $\hat j_{i}$ is not anymore odd under space reflection around the link between sites $i$ and $i+1$, so that in the stationary state a non-zero current can flow even if the density profile is homogeneous (and even under the previous space-inversion transformation), which is quite intuitive given the explicit breaking of inversion symmetry in this problem.

We employ the MPDO method to analyze the steady-state properties 
of a system with size  $L=22$ initialized in the ground state of the free Hamiltonian 
$\hat {\mathcal H}_{0}$ for $N=11$ and subject to such dissipation.
The results in the inset of Fig.~\ref{excitation.perturbation.L22} show that the steady state 
is not homogeneous and that a relatively high degree of inhomogeneity 
$\frac{\langle \hat n_{L} \rangle - \langle \hat n_{1} \rangle }{\langle \hat n_{L/2} \rangle} \approx 1$ 
is found also for small perturbations $\epsilon = 0.05$. 
This is not to be confused with the phase-separation instability which characterizes 
the ferromagnetic parent Hamiltonian $\hat {\mathcal H}'_{p, \mathrm{spin}}$. 
Indeed, if PBC are considered, the system becomes homogeneous and a current starts flowing in it 
(not shown here).

P-wave superconducting correlations are affected by such inhomogeneity.
Whereas for $\epsilon=0$ the correlations $\big[ G^{(p)}_{j,l} \big]_{\rm ss}$ 
do not show a significant dependence on $|j-l|$, this is not true even 
for small perturbations $\epsilon \leq 0.05$.
In order to remove the effect of the inhomogeneous density,
in Fig.~\ref{excitation.perturbation.L22} we show the value of properly rescaled p-wave correlations:
\begin{equation}\label{eq:gprime}
  \big[ G'^{(p)}_{j,l} \big]_{\rm ss} \equiv \langle \, \hat O'^{(p) \dagger}_j \, \hat O'^{(p)}_l \, \rangle_{\rm ss} = 
  \frac{(N/L)^{4}\, \langle \hat O^{(p) \dagger}_j \hat O^{(p)}_l \rangle_{\rm ss}}
  {\langle \hat n_j \rangle_{\rm ss} \langle \hat n_{j+1} \rangle_{\rm ss} \langle \hat n_l \rangle_{\rm ss} \langle \hat n_{l+1} \rangle_{\rm ss} }   
\end{equation}
where $\hat O'^{(p)}_j =(N/L)^2 \hat O^{(p)}_j / (\langle \hat n_j \rangle_{\rm ss} \langle \hat n_{j+1} \rangle_{\rm ss})$.
An exponential decay behavior appears as a function of $|j-l|$, which becomes more pronounced when $\epsilon$ is increased.
Even if the simulation is performed on a finite short system, for significant perturbations, $\epsilon = 0.1$, the value of $\big[ G'^{(p)}_{j,l} \big]_{\rm ss}$ decays of almost two decades, so that the exponential behavior is  identified with reasonable certainty.

In Appendix~\ref{app:ParentHamiltonian} we discuss some interesting analogies of these results 
with the properties of the ground state of the parent Hamiltonian 
$\hat {\mathcal H}'_{p, \epsilon} = J \sum_j \hat L_{j, \epsilon}'^\dagger \hat L'_{j, \epsilon}$. 
It should be stressed that, since $\hat {\mathcal H}'_{p, \epsilon}$ does not have a zero-energy 
ground state, there is no exact correspondence between its ground state 
and the steady states of $\mathcal L_{\epsilon}'$.

Concluding, we mention that a similar analysis can be done introducing an analogous perturbation 
in the operator $\hat C_j^\dagger$; our study did not observe any qualitative difference (not shown).

\subsubsection{Perturbations due to unitary dynamics}

\begin{figure}
  \centering
  \includegraphics[scale=0.4]{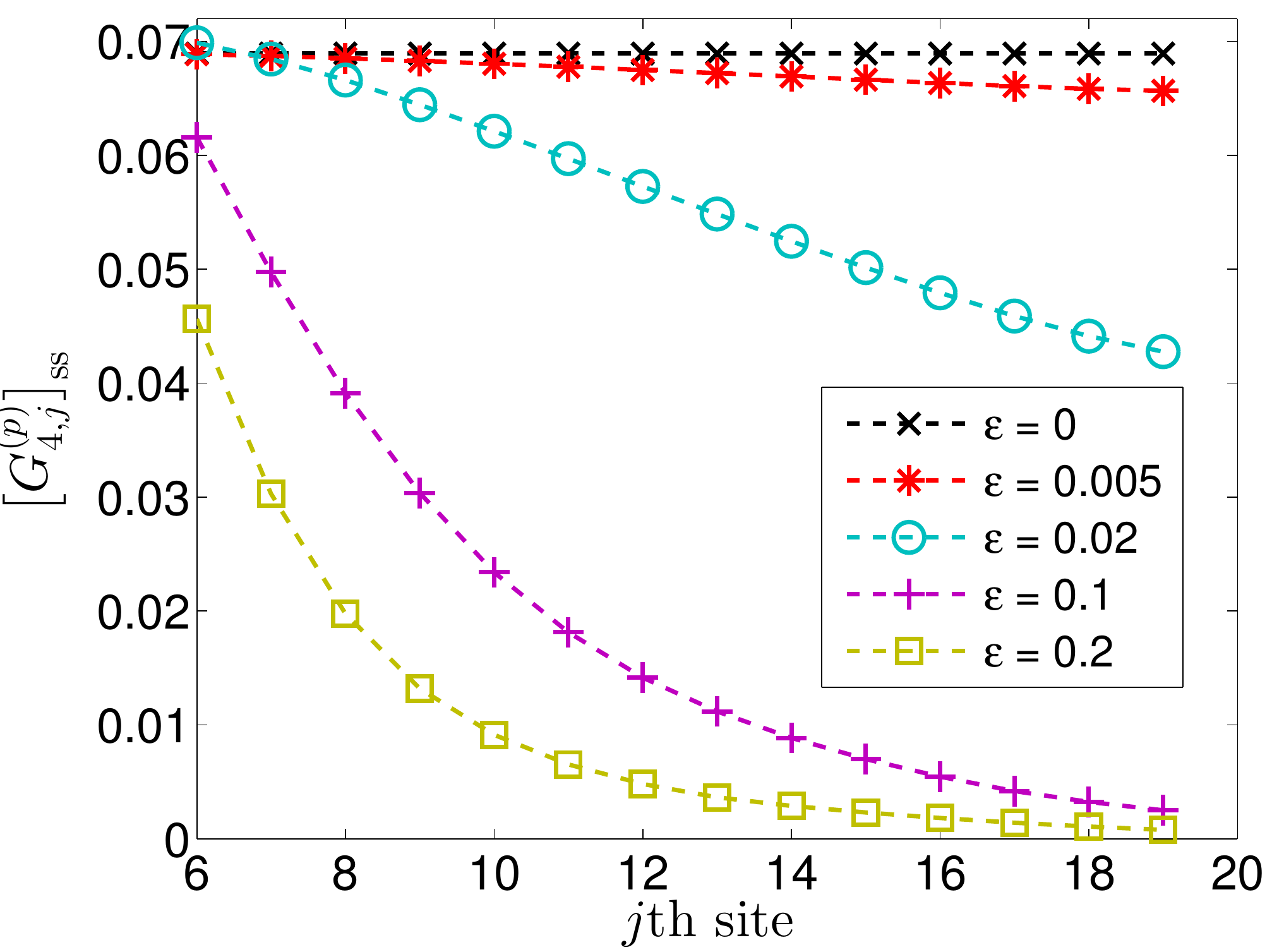} 
  \includegraphics[scale=0.4]{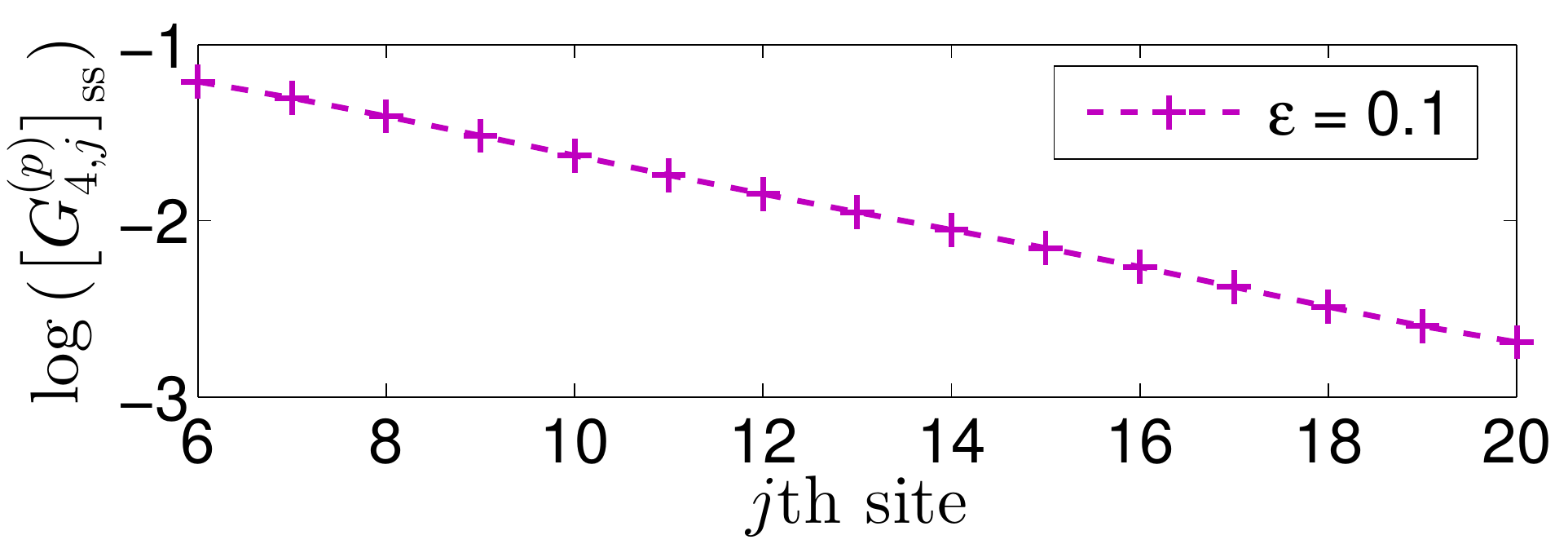}
  \caption{(Color online) (Top) Pairing correlations 
    $\big[ G^{(p)}_{4,j} \big]_{\rm ss}$ for the steady state of the dynamics 
    in the presence of a Hamiltonian perturbation~\eqref{eq:Lind:pert:Ham}. 
    The calculation of the steady state is performed with MPDO technique for $L=22$ and $N=11$. 
    (Bottom) The decay of $\big[ G^{(p)}_{4,j} \big]_{\rm ss}$ 
    is exponential in $j$ (here, $\epsilon=0.1$).
  }
  \label{free.fermion.perturbation}
\end{figure}

An alternative way of perturbing the dynamics of $\mathcal L'$ in Eq.~\eqref{eq:master:equation:N:1} 
is to introduce a Hamiltonian into the system, chosen for simplicity to be the already-introduced 
free Hamiltonian $\hat {\mathcal H}_0$:
\begin{equation}
  \frac{\partial}{\partial t}\hat \rho =
  -i [ \epsilon \hat {\mathcal H}_0, \hat \rho] + \mathcal L[\hat \rho].
  \label{eq:Lind:pert:Ham}
\end{equation}
Using the MPDO method to characterize the steady state of the dynamics, we analyze 
the spatial decay of the pairing correlations for $L=22$ and at half-filling ($N=11$); 
the initial state is set in the same way as in the previous section. 
In Fig.~\ref{free.fermion.perturbation} (top) we display the results:
even for very small perturbations the pairing correlator 
$\big[ G^{(p)}_{4,j} \big]_{\rm ss}$ decays rapidly in space. 
The long-distance saturation observed in the absence of perturbations is lost 
and qualitatively different from this result.
In Fig.~\ref{free.fermion.perturbation} (bottom) we highlight that the decay is exponential.

Summarizing, in all the cases that we have considered, the p-wave pairing correlations of the stationary state $\big[ G^{(p)}_{j,l} \big]_{\rm ss}$ are observed to decay as a function of $|j-l|$. Due to the interplay between the targeted dissipative dynamics and the perturbations, which do not support a p-wave ordered dark state, the steady state is mixed, similar to a finite temperature state. From this intuition, the result is easily rationalized: Any (quasi) long range order is destroyed in one-dimensional systems at finite temperature. We note that the true long range order found in the unperturbed case (correlators saturating at large distance; opposed to the more generic quasi-long range order defined with algebraic decay) is non-generic in one-dimensional systems and a special feature of our model, see \cite{Iemini_2015} for a thorough discussion. However, the destruction of any such order via effective finite temperature effects must be expected on general grounds. The absence of quasi-long-range p-wave superconducting order, which in one-dimension only occurs at zero-temperature for pure state, is likely to be in connection with this fact.

\subsubsection{Perturbation strength}

\begin{figure}
  \centering
  \includegraphics[scale=0.45]{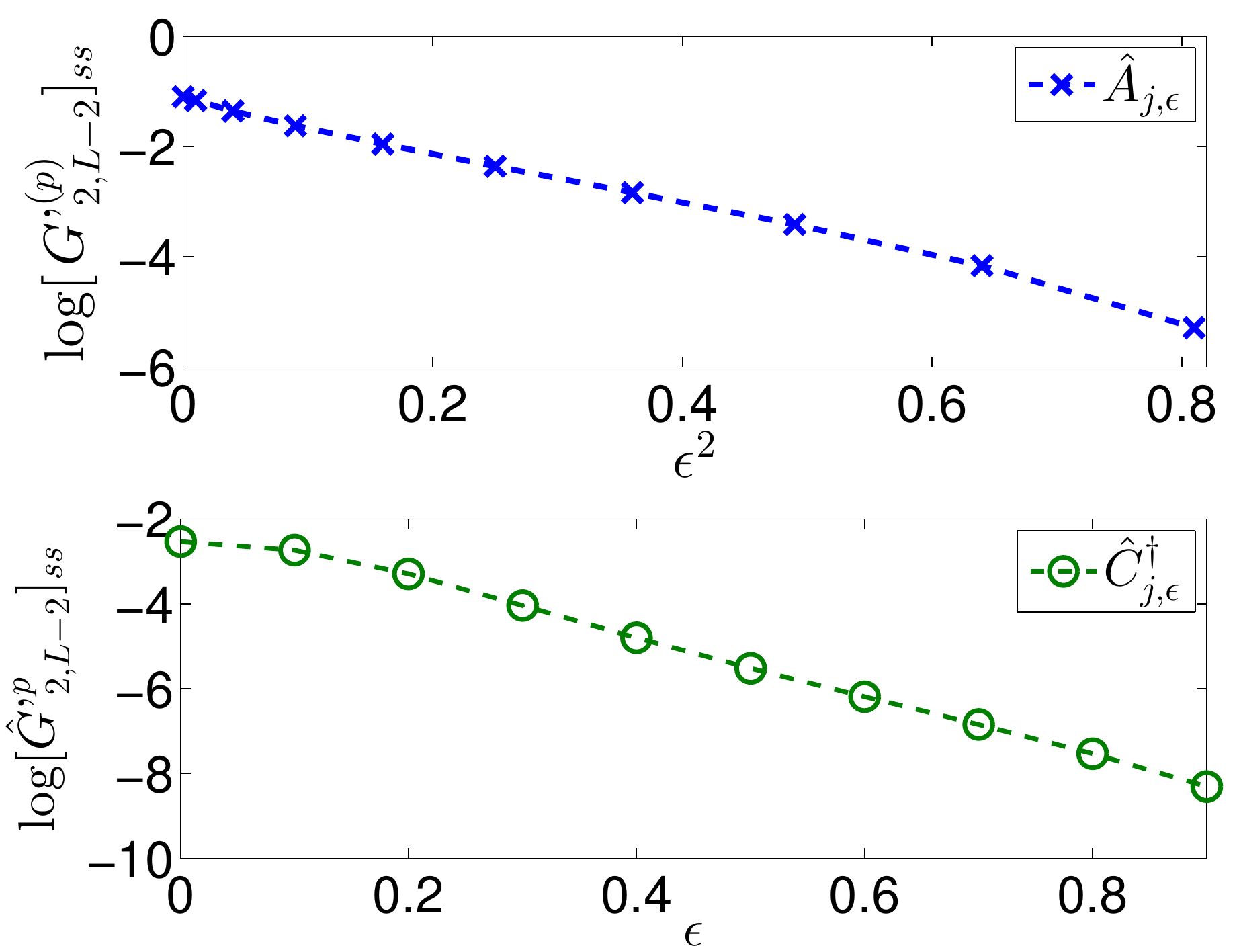}
  \caption{  (Color online) $\big[ G'^{(p)}_{2,L-2} \big]_{\rm ss}$ in the presence of a perturbed Lindblad operator as a function of the perturbation strength $\epsilon$. 
  The perturbation is considered both for the $\hat A_j$ (top) and $\hat C^\dagger_j$ (bottom)  operators (see text for the definitions). 
  The calculation is done with RK integration of the equation of motion for $L=8$ and $N=4$.
  }
  \label{hamiltonian.lindblad.perturb}
\end{figure}

Finally, we perform a quantitative investigation of the dependence of the pairing correlations on the perturbation strength, $\epsilon$.

{\it{Lindblad perturbation}} -- In Fig.~\ref{hamiltonian.lindblad.perturb}  we plot the p-wave superconducting 
correlation $\big[ G^{(p)}_{2,L-2} \big]_{\rm ss}$ 
of a system of length $L=8$ as a function of the intensity of the perturbation 
$\epsilon$ in $\hat L'_{j, \epsilon}$ (for completeness, the complementary case
$\hat L'^{(2)}_{j,\epsilon} = \hat C_{j,\epsilon}^\dagger \hat A_{j}$, with
$\hat C_{j,\epsilon}^\dagger = \hat a_j^\dagger+(1-\epsilon)\hat a^\dagger_{j+1}$, is also included). 
Our data confirm that correlations undergo a clear suppression in the presence of $\epsilon \neq 0$, which in one case is exponential in $\epsilon$ and in the other in $\epsilon^2$.
The calculation is performed through RK integration of the dynamics. 


\begin{figure}
  \centering
  \includegraphics[scale=0.45]{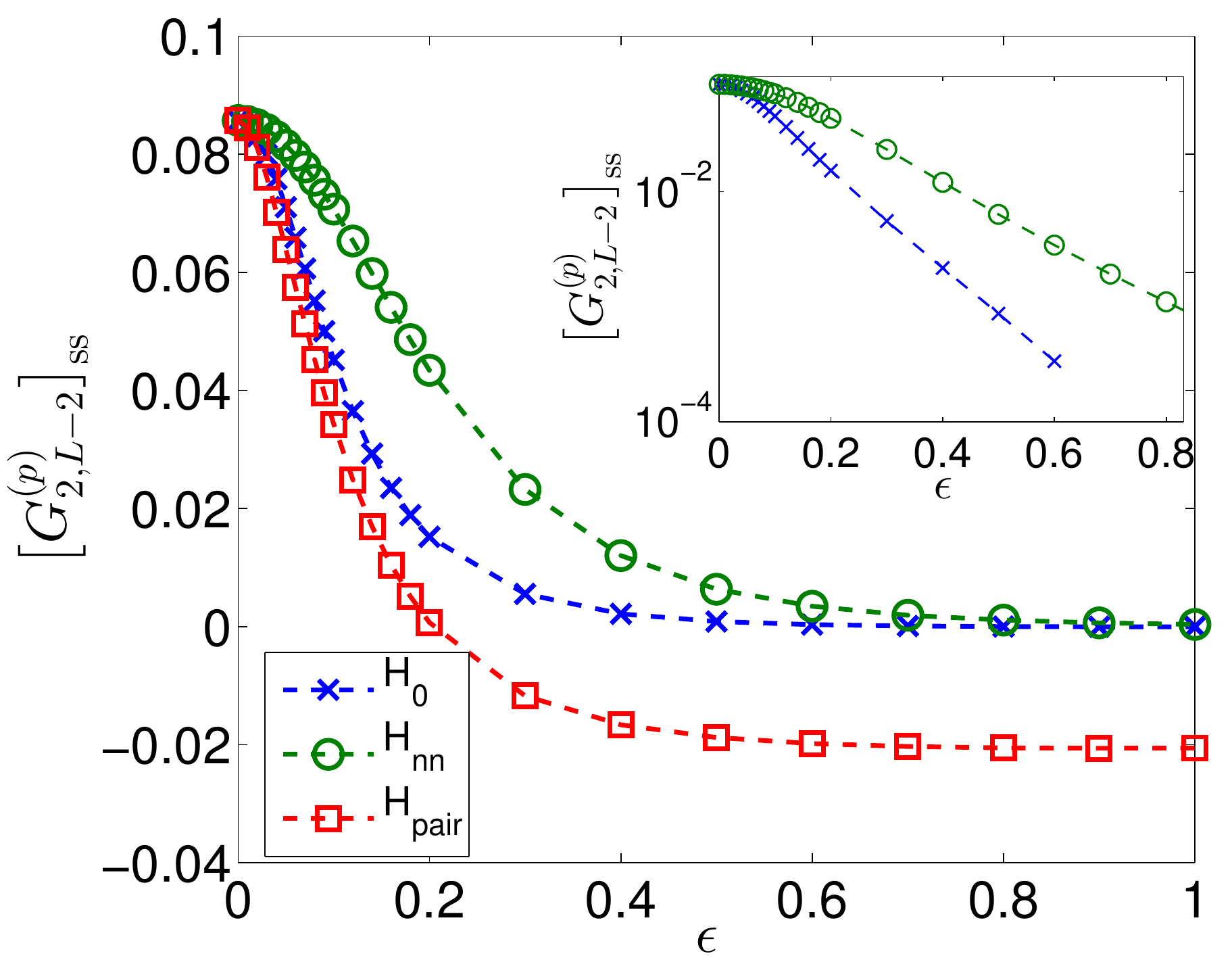}
  \caption{
  (Color online) $\big[ G^{(p)}_{2,L-2} \big]_{\rm ss}$ 
    in the presence of a perturbing Hamiltonian as a function of the perturbation strength $\epsilon$. 
    We consider $\hat \Ham_0$, $\hat \Ham_{\rm nn}$ and $\hat \Ham_{\rm pair}$ (see text for the definitions). The inset highlights the exponential decay with $\epsilon$.
      }
  \label{hamiltonian.lindblad.perturb.2}
\end{figure}


{\it{Hamiltonian perturbation}} -- We begin with the two cases:
$\hat {\mathcal H}_{0}$ and $\hat {\mathcal H}_{\rm nn} = - J \sum_j \hat n_j \hat n_{j+1}$.
Fig.~\ref{hamiltonian.lindblad.perturb.2} shows, in both cases, an exponential decay 
to zero of $\big[ G'^{(p)}_{2,L-2} \big]_{\rm ss}$ when $\epsilon$ is increased.
On the contrary, a Hamiltonian which introduces p-wave correlations in the system, such as
\begin{equation}
  \hat {\mathcal H}_{\rm pair} = -J \sum_{j,l} (\hat a^\dagger_{j} \hat a^\dagger_{j+1} \hat a_{l+1} \hat a_l + {\rm H.c.}),
  \label{pairing.hamiltonian:2}
\end{equation}
changes the value and the sign of $\big[ G'^{(p)}_{2,L-2} \big]_{\rm ss}$, leaving it different from zero.

Concluding, we have shown that in all the considered cases, perturbations of both dissipative 
and Hamiltonian form are detrimental to the creation of a p-wave superconductor.
 This is rationalized by the mixedness of the stationary state in that case, and parallels a finite temperature situation.
In any generic, algebraically ordered system at T=0, one has gapless modes.

\section{Two wires}\label{sec:N:2}

An intuitive explanation of why the dissipative setup discussed in Sec.~\ref{sec:N:1} does not show topological dark states with fixed number of particles is the fact that this constraint fixes the parity of the state, and thus no topological degeneracy can occur. 
It has already been realized in several works that a setup with two parallel wires can overcome 
this issue~\cite{Fidkowski_2011, Sau_2011, Cheng_2011, Kraus_2013, Ortiz_2014, Iemini_2015, Lang_2015}. 
In this case it is possible to envision a number-conserving p-wave superconducting Hamiltonian 
which conserves the parity of the number of fermions in each wire: such symmetry can play the role 
of the parity of the number of fermions for $\hat {\mathcal H}_{\rm K}$ in Eq.~\eqref{eq:Kitaev:Hamiltonian}.
Several equilibrium models have already been discussed in this context; here we consider 
the novel possibility of engineering a topological number-conserving p-wave superconductor with Markovian dynamics.

\subsection{Steady states}

Let us study a system composed of two wires with spinless fermions described by the canonical 
fermionic operators $\hat a_j^{(\dagger)}$ and $\hat b_j^{(\dagger)}$.
For this model we consider three kinds of Lindblad operators:
\begin{subequations}
  \label{eq:Lind:N:2}
  \begin{align}
    & \hat L''_{a,j} = \hat C_{a,j}^\dagger \hat A_{a,j}; \label{eq:Lind_1-wire}
    \\
    & \hat L''_{b,j} = \hat C_{b,j}^\dagger \hat A_{b,j}; \label{eq:Lind_2-wire}
    \\
    & \hat L''_{I,j} = \hat C_{a,j}^\dagger \hat A_{b,j} + \hat C^\dagger_{b,j}\hat A_{a,j}.
    \label{eq:Lind_Int}
  \end{align}
\end{subequations}
We now characterize the dark states of the Markovian dynamics induced by these 
operators for a two-leg ladder of length $L$ with hard-wall boundary conditions:
\begin{equation}
  \frac{\partial}{\partial t}\hat \rho =
  \mathcal L''[\hat \rho] = \gamma
  \sum_{j=1}^{L-1} \sum_{\Lambda= a,b,I} \left[ \hat L_{\Lambda,j}'' \hat \rho \hat L_{\Lambda,j}''^\dagger - \frac 12 \{ \hat L_{\Lambda,j}''^\dagger \hat L''_{\Lambda,j}, \hat \rho \} \right].
  \label{eq:master:equation:N:2}
\end{equation}
In particular, we will show that, for every fermionic density different from the completely empty and
filled cases, there are always two steady states.

It is easy to identify the linear space $\mathcal S_N$ of states which are annihilated by the $\hat L''_{a,j}$ and $\hat L''_{b,j}$ and have a total number of particles $N$:
\begin{equation}
  \mathcal S_N = \text{span} \{ 
  \ket{\psi_{a,0}} \hspace{-0.05cm} \ket{\psi_{b,N}},
  \ket{\psi_{a,1}} \hspace{-0.05cm} \ket{\psi_{b,N-1}},
  \ldots,
  \ket{\psi_{a,N}} \hspace{-0.05cm} \ket{\psi_{b,0}}
  \}.
\end{equation}
where the states $\ket{\psi_{\alpha,N}}$ are those defined in Eq.~\eqref{eq:proj:N:stst} 
for the wire $\alpha=a,b$. Let us consider a generic state in $\mathcal S_N$:
\begin{equation}
  \ket{\psi} = \sum_{m = 0}^{N} \alpha_{m} 
  \ket{\psi_{a,m}} \ket{\psi_{b,N-m}},
  \quad \sum_{m = 0}^{N} |\alpha_{m}|^2 =1.
\end{equation}
From the condition $\hat C^\dagger_j \ket{\psi_\sigma} = - \hat A_j \ket{\psi_\sigma}$ we obtain:
\begin{subequations}
  \begin{align}
    \hat C^\dagger_j \ket{\psi_{N-1}} &= - \hat A_j \ket{\psi_{N+1}}, \quad N \in (0,2L) \\
    0 &= - \hat A_j \ket{\psi_{1}} , \\
    \hat C^\dagger_j \ket{\psi_{2L-1}} &= 0 ,
  \end{align}
\end{subequations}
and when we impose the condition $\hat L''_{I,j} \ket \psi = 0$:
\begin{widetext}
  \begin{align}
    \hat L''_{I,j} \ket{\psi} 
    =& \sum_{m = 0}^{N-1} \alpha_{m} 
    \hat C^{\dagger}_{a,j}\hat A_{b,j} \ket{\psi_{a,m}} \ket{\psi_{b,N-m}} + 
    \sum_{m = 1}^{N} \alpha_{m} \hat C^\dagger_{b,j}\hat A_{a,j} \ket{\psi_{a,m}} \ket{\psi_{b,N-m}}
    = \nonumber \\
    =& \sum_{m = 0}^{N-1} \alpha_{m} 
    \hat C^{\dagger}_{a,j}\hat A_{b,j} \ket{\psi_{a,m}} \ket{\psi_{b,N-m}} - \sum_{m = 2}^{N+1} \alpha_{m} \hat C^\dagger_{a,j}\hat A_{b,j} \ket{\psi_{a,m-2}} \ket{\psi_{b,N-m+2}}=0.
  \end{align}
\end{widetext}
The result is $\alpha_m = \alpha_{m+2}$, so that two linearly independent states 
can be constructed which are annihilated by all the Lindblad operators in~\eqref{eq:Lind:N:2}:
\begin{subequations}
  \label{eq:states:ee}
  \begin{align}
    \ket{\psi_{N,ee}} =& \frac{1}{\mathcal N_{N,ee}^{1/2}}\sum_m \ket{\psi_{a,2m}}\ket{\psi_{b,N-2m}}, \\
    \ket{\psi_{N,oo}} =& \frac{1}{\mathcal N_{N,oo}^{1/2}}\sum_m \ket{\psi_{a,2m-1}}\ket{\psi_{b,N-2m+1}}. 
  \end{align}
\end{subequations}
The subscripts $ee$ and $oo$ refer to the fermionic parities in the first 
and second wire assuming that $N$ is even; $\mathcal N_{N,ee}$ and $\mathcal N_{N,oo}$
are normalization constants~\cite{Iemini_2015}. For $N$ odd one can similarly construct 
the states $\ket{\psi_{N,eo}}$ and $\ket{\psi_{N,oe}}$.
By construction, the states that we have just identified are the only dark states of the dynamics.

It is an interesting fact that at least two parent Hamiltonians are known 
for the states in~\eqref{eq:states:ee}, as discussed in Refs.~\cite{Iemini_2015, Lang_2015}. 
We refer the reader interested in the full characterization of the topological properties 
of these steady-states to those articles.

Finally, let us mention that the form of the Lindblad operators in~\eqref{eq:Lind:N:2}
is not uniquely defined. For example one could replace $\hat L''_{I,j}$ in Eq.~\eqref{eq:Lind_Int}
with the following: 
\begin{equation}
  \hat L''_{I,j} = \left( \hat C_{a,j}^\dagger + \hat C_{b,j}^\dagger \right) \left( \hat A_{a,j} + \hat A_{b,j} \right),
  \label{eq:Lind_Int2}
\end{equation}
without affecting the results~\cite{Iemini_2015}.
 The latter operator is most realistic for an experimental implementation, as we point out below.

\subsection{P-wave superconductivity}

Let us now check that the obtained states are p-wave superconductors.
Similarly to the single-wire protocol discussed in Eq.~\eqref{eq:order:corr},
the explicit calculation~\cite{Iemini_2015} shows that p-wave correlations 
saturate to a final value at large distances in the thermodynamic limit 
[for the two-leg ladder we consider $\nu=N/(2L)$]
\begin{equation}
  \bra{\psi_{N,ee}} \hat O_j^{(p)\dagger} \hat O^{(p)}_l \ket{\psi_{N,ee}} 
  \xrightarrow{|j-l| \rightarrow \infty} \nu^2(1-\nu)^2.
\end{equation}
This relation clearly highlights the p-wave superconducting nature of the states.

\subsection{Dissipative gap}

In order to demonstrate that the asymptotic decay rate $\lambda_{\rm ADR}$ associated 
to $\mathcal L''$ tends to $0$ in the thermodynamic limit, we consider the parent Hamiltonian of the model:
\begin{align}
  \label{eq:hamiltonian}
  \hat {\mathcal H}_p'' \! = \! & - 4 J \!\! \sum_{\substack{j=1 \\ \alpha =a,b}}^{L-1} \!\! \Big[ (\hat \alpha^\dagger_{j}\hat \alpha_{j+1} \! + \! \text{H.c.}) - \! (\hat n_j^{\alpha} + \hat n_{j+1}^{\alpha}) + \! \hat n_j^{\alpha} \hat n_{j+1}^{\alpha} \Big] \nonumber \\
  & -2 J \sum_{j=1}^{L-1} \Big[ (\hat n_j^a + \hat n_{j+1}^a)(\hat n_j^b + \hat n_{j+1}^b) - (\hat a^\dagger_{j}\hat a_{j+1} \hat b^\dagger_{j} \hat b_{j+1} \nonumber\\ 
    & \hspace{0.65cm} + \hat a_{j}^\dagger \hat a_{j+1} \hat b^\dagger_{j+1} \hat b_{j} - 2 \hat b^\dagger_{j} \hat b^\dagger_{j+1} \hat a_{j+1} \hat a_{j} + {\rm H.c.}) 
    \Big] ,
\end{align}
where $J>0$ is a typical energy scale setting the units of measurement.
This Hamiltonian has been extensively analyzed in Ref.~\cite{Iemini_2015}.
Numerical simulations performed with the density-matrix renormalization-group algorithm 
assess that $\hat {\mathcal H}_p''$ is gapless and that the gap is closing as $1/L^2$.
According to the discussion in Sec.~\ref{sec:I:A}, the asymptotic decay rate $\lambda_{\rm ADR}$ 
associated to the Lindbladian $\mathcal L''$ closes in the thermodynamic limit with a scaling 
which is equal to $\sim L^{-2}$ or faster. 
This is true both for periodic and hard-wall boundary conditions.

\subsection{Experimental implementation}

The Lindblad operators in Eqs.~\eqref{eq:Lind_1-wire}, \eqref{eq:Lind_2-wire} and~\eqref{eq:Lind_Int2}
lend themselves to a natural experimental implementation.
The engineering of terms like $\hat L''_{a,j}$ and $\hat L''_{b,j}$ has been extensively discussed in Ref.~\cite{Diehl_2011} starting from ideas originally presented in Ref.~\cite{Diehl_2008}. As we will see, the Lindblad operator $\hat L''_{I,j}$ in Eq.~\eqref{eq:Lind_Int2} is just a simple generalization.

The idea is as follows: a superlattice is imposed which introduces in the system additional 
higher-energy auxiliary sites located in the middle of each square of the lower sites target lattice. 
Driving lasers are then applied to the system, whose phases are chosen such that the excitation 
to the auxiliary sites happens only for states $\ket{\varphi}$ such that
$(\hat A_{a,j} + \hat A_{b,j}) \ket{\varphi} \neq 0$.
If the whole system is immersed into, e.g., a Bose-Einstein condensate reservoir, atoms located 
in the auxiliary sites can decay to the original wire by emission 
of a Bogoliubov phonon of the condensate.
This process is isotropic and, for a wavelength of the emitted phonons comparable to the lattice spacing, gives rise to the four-site creation part 
with relative plus sign: $\hat C^\dagger_{a,j} + \hat C^\dagger_{b,j}$. 

\subsection{Perturbations}

An important property of topological Hamiltonians is the robustness of their edge physics to local perturbations. Similar features have been highlighted in the case of topological superconductors where the setup is not number conserving~\cite{Diehl_2011, Bardyn_2013}. The goal of this section is to probe the resilience of the twofold-degenerate steady states of $\mathcal L''$. A conclusive analysis is beyond our current numerical possibilities; here we present some preliminary results obtained via exact diagonalization methods.

We consider the natural choice of Lindblad operators Eqs.~(\ref{eq:Lind_1-wire},\ref{eq:Lind_2-wire},\ref{eq:Lind_Int2}), subject to perturbations:
\begin{subequations}
\label{eq:pert:exc:2}
\begin{align}
&  \hat L''_{a,j,\epsilon} = \hat C_{a,j}^\dagger \hat A_{a,j,\epsilon}; 
  \quad
  \hat A_{a,j,\epsilon} = \hat a_j-(1-\epsilon)\hat a_{j+1};
   \\
&  \hat L''_{b,j,\epsilon} = \hat C_{b,j}^\dagger \hat A_{b,j,\epsilon}; 
  \quad
  \hat A_{b,j,\epsilon} = \hat b_j-(1-\epsilon)\hat b_{j+1}; 
  \\
& \hat L''_{I,j} = \left( \hat C_{a,j}^\dagger + \hat C_{b,j}^\dagger \right) \left( \hat A_{a,j,\epsilon} + \hat A_{b,j,\epsilon} \right) ; \quad \epsilon \in \mathbb R
\end{align}
\end{subequations}
Those define a perturbed Lindbladian $\mathcal L''_{\epsilon}$. They are a simple generalization of those defined in Eq.~\eqref{eq:pert:exc} for the single-wire setup.

\begin{figure}[t]
  \centering
  \includegraphics[scale=0.45]{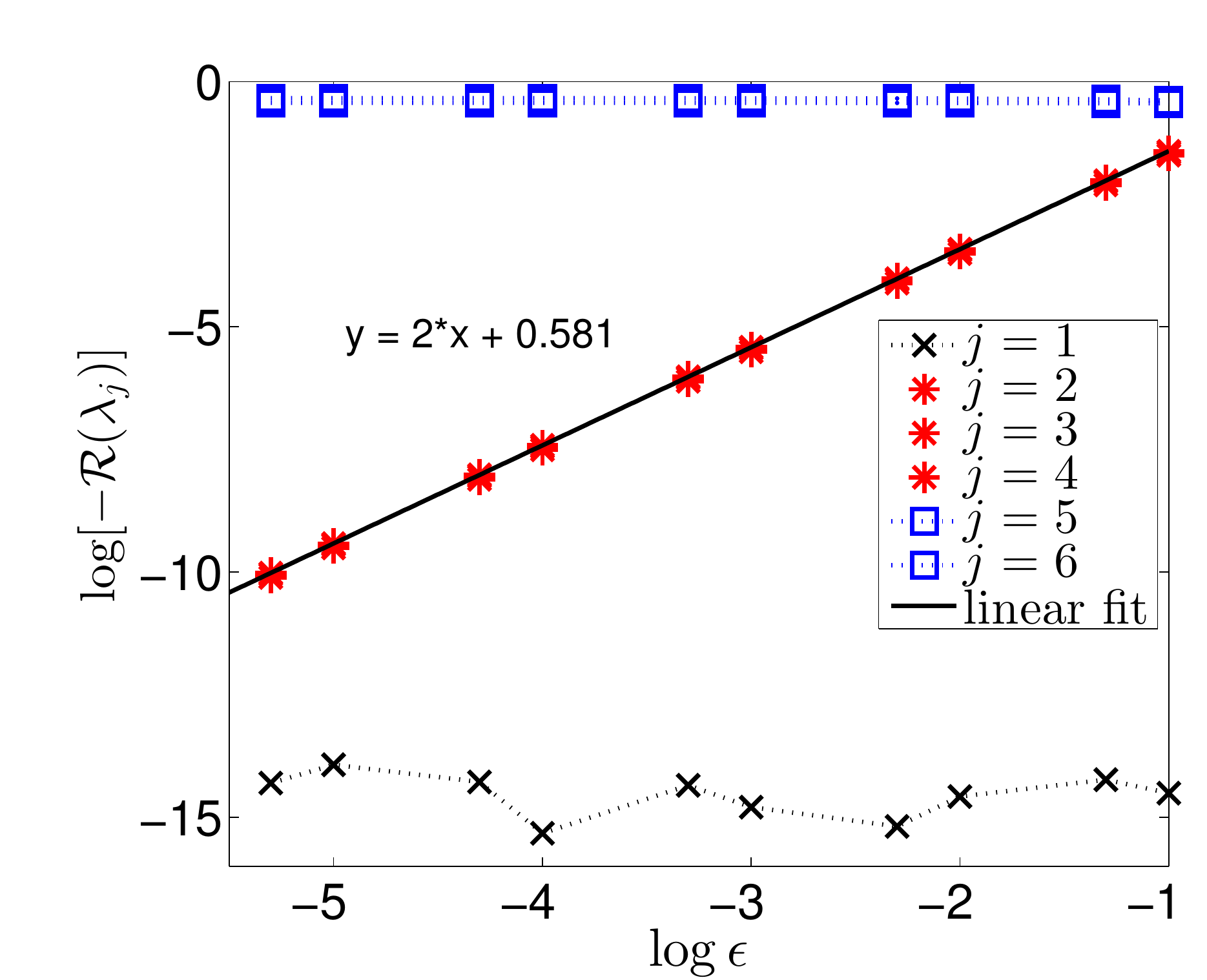}
  \caption{  (Color online) Real part of the first six eigenvalues of the Lindbladian operator $\mathcal L''_{\epsilon}$ for $L=6$ and $N=6$ as a function of $\epsilon$. Eigenvalues $\lambda_j$ are sorted according to increasing $- \Re (\lambda_j)$. The plot highlights the presence of a $\lambda=0$ eigenvalue (within numerical accuracy $10^{-15}$), of three eigenvalues which scale as $\epsilon^{2}$ and of other eigenvalues of magnitude $\sim 1$. }
  \label{two.wires.perturbation.L6N6}
\end{figure}

Let us begin our analysis by showing that for small sizes $L \sim 6$ the degeneracy of the steady space for $\epsilon=0$ is broken. 
Let us first remark that for $\epsilon=0$ the steady space is four-fold degenerate; a possible parameterization is:
\begin{align}
 \mathcal B =  \{ & \ket{\psi_{N,ee}} \hspace{-0.05cm} \bra{\psi_{N,ee}} ,
 \quad
 \ket{\psi_{N,ee}} \hspace{-0.05cm} \bra{\psi_{N,oo}} ,\\
 &\ket{\psi_{N,oo}} \hspace{-0.05cm} \bra{\psi_{N,ee}} ,
 \quad
 \ket{\psi_{N,oo}} \hspace{-0.05cm} \bra{\psi_{N,oo}} \}.
\end{align}
A direct inspection of the eigenvalues of $\mathcal L_{\epsilon}$ shows that this degeneracy is broken once $\epsilon \neq 0$. Results, shown in Fig.~\ref{two.wires.perturbation.L6N6} for a fixed lattice size $L=6$ and $N=6$, display a quadratic splitting of the steady steady degeneracy with the perturbation strength.

\begin{figure}[t]
  \centering
  \includegraphics[scale=0.45]{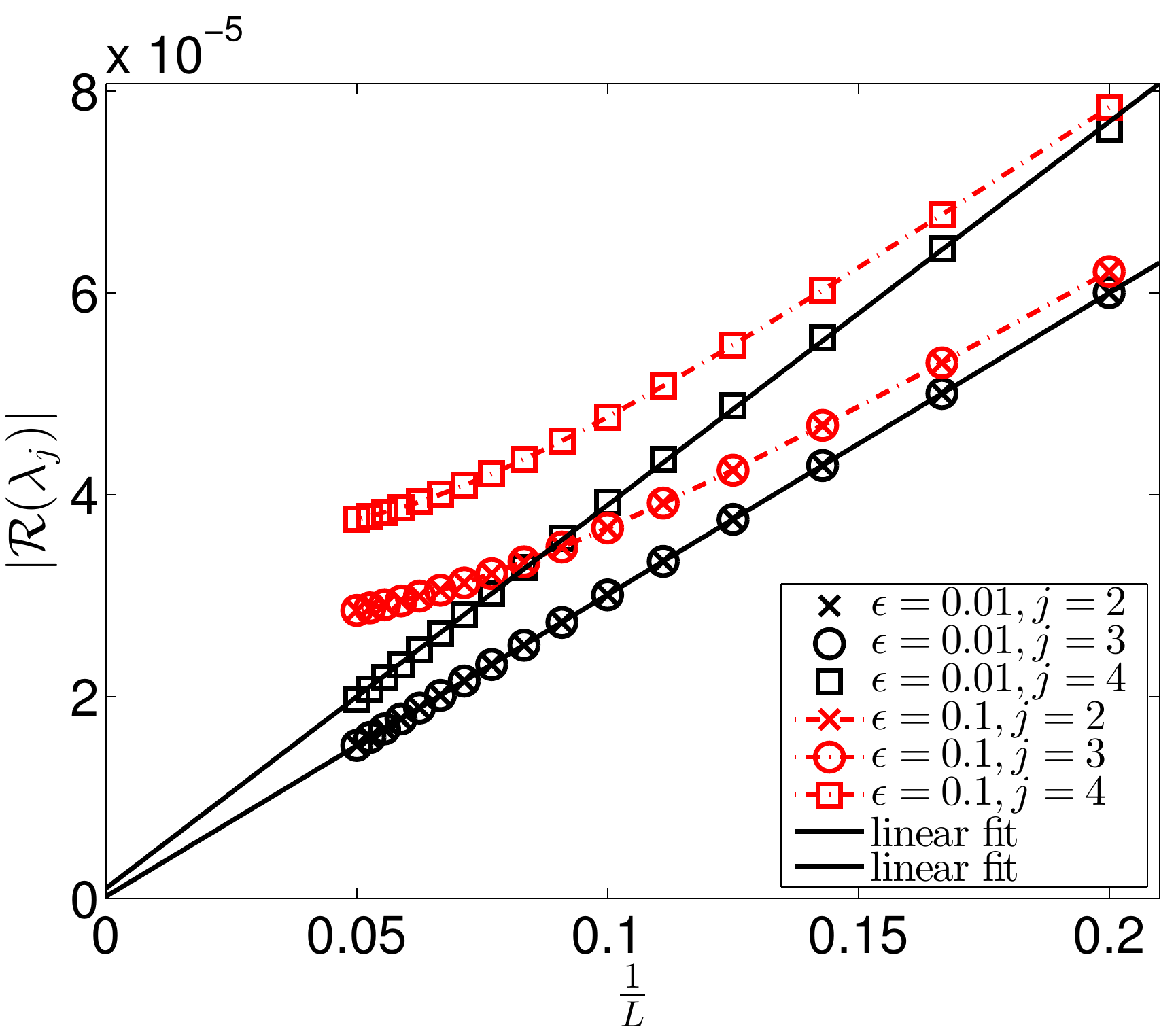}
  \includegraphics[scale=0.45]{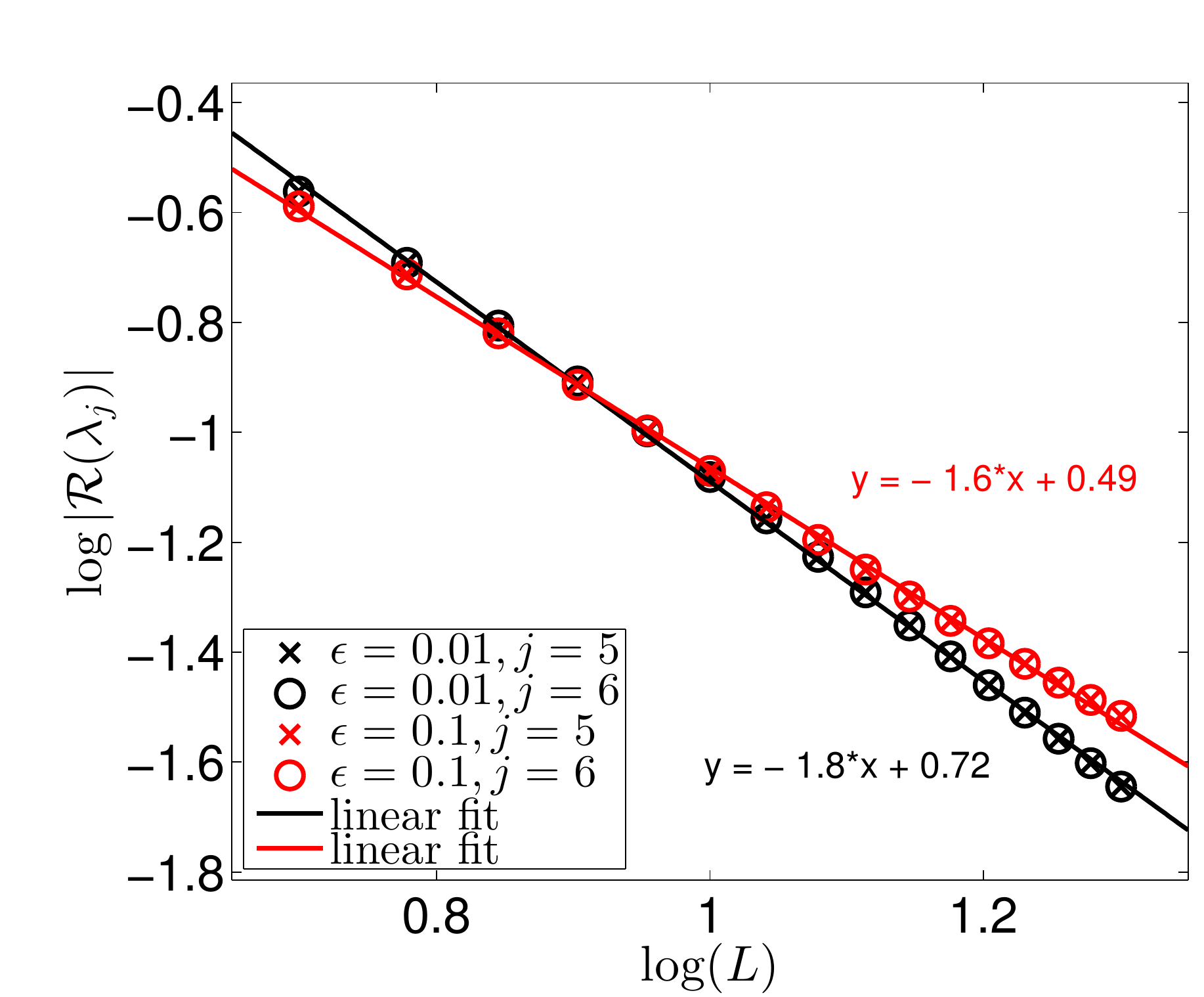}
  \caption{(Color online) Real part of the eigenvalues $j=2$, $3$ and $4$ (top) and $j=5$ and $6$ (bottom) of the Lindbladian operator $\mathcal L''_{\epsilon}$ for $N=2$ as a function of $L$ (here, $L\leq20$).
  The two values $\epsilon=0.1$ and $\epsilon=0.01$ are considered.
  In the top panel, the values of the eigenvalues relative to $\epsilon=0.1$ have been rescaled by $0.01$ in order to facilitate the readability of the plot.}
  \label{two.wires.perturbation.fixedN2}
\end{figure}

Let us now check the behavior with the system size of the first eigenvalues of the system for longer system sizes. 
In order to obtain a reasonable number of data, the extreme choice of setting $N=2$ in all simulations has been taken, which allows us to analyze system sizes up to $L=20$. Results shown in Fig.~\ref{two.wires.perturbation.fixedN2} (top) show that the Liouvillian eigenvalues related to the steady-state degeneracy display an algebraic scaling $\lambda_{\rm ADR}\sim L^{-1}$ in the accessible regime of system sizes for small perturbations ($\epsilon=10^{-2}$), while they are gapped for larger perturbations ($\epsilon=10^{-1}$). 
Note that, for the system sizes which could be accessed, larger eigenvalues clear display an algebraic decay, as shown in Fig.~\ref{two.wires.perturbation.fixedN2} (bottom), also for $\epsilon=0.1$. The scaling of the eigenvalues related to the steady state degeneracy is not exponential and thus in principle should not be connected to the topological properties of the system. However, these preliminary considerations suffer from two significant biases: (i) the small considered sizes, (ii) the fact that they are not performed at exactly fixed density, and (iii) the very low filling. A more thorough analysis is left for future work.

\section{Conclusions}\label{sec:conc}

In this article we have discussed the dissipative quantum state preparation of a p-wave superconductor in one-dimensional fermionic systems with fixed number of particles. 
In particular, we have presented two protocols which have been fully characterized in the presence 
of hard-wall boundaries. Whereas the former does not display topological property, 
the latter features a two-dimensional steady space to be understood in terms of boundary 
Majorana modes for any number of fermions. Through the analysis of a related parent Hamiltonian, 
we are able to make precise statements about the gapless nature of the Lindbladian 
super-operators associated to both dynamics.

The peculiar form of the master equations considered in this article allows for the exact 
characterization of several properties of the system, and in particular of the steady state, 
even if the dynamics is not solvable with the methods of fermionic linear 
optics~\cite{Prosen_2008, Bravyi_2012} exploited in Refs.~\cite{Diehl_2011, Bardyn_2013}. 
This result is very interesting \textit{per se}, as such examples are usually rare 
but can drive physical intuition into regimes inaccessible without approximations. 
It is a remarkable challenge to investigate which of the properties presented so far are general 
and survive to modifications of the environment, and which ones are peculiar of this setup.

Using several numerical methods for the study of dissipative many-body systems, we have presented a detailed analysis of the robustness to perturbations of these setups. Through the calculation of the proper p-wave correlations we have discussed how external perturbations can modify the nature of the steady state. In the ladder setup, where the steady states are topological, we have presented preliminary results on the stability of the degenerate steady-space of the system.

The analysis presented here has greatly benefited from exact mathematical relations 
between the properties of the Lindbladian and of a related parent Hamiltonian. 
Since the study of closed systems is much more developed than that of open systems 
both from the analytical and from the numerical points of view, a more detailed understanding of the relations between Lindbladians and associated parent Hamiltonian operators 
stands as a priority research program.

\acknowledgments

We acknowledge enlightening discussions with C. Bardyn, G. De Palma, M. Ippoliti and A. Mari.
F. I. acknowledges financial support by the Brazilian agencies FAPEMIG, CNPq, and
INCT- IQ (National Institute of Science and Technology for Quantum Information). 
D. R. and L. M. acknowledge the Italian MIUR through FIRB Project No. RBFR12NLNA.
R. F. acknowledges financial support from the EU projects SIQS and QUIC and from Italian MIUR via PRIN Project No. 2010LLKJBX.
S. D. acknowledges support via the START Grant No. Y 581-N16, the German Research Foundation through ZUK 64, and through  the  Institutional
Strategy of the University of Cologne within the German Excellence Initiative (ZUK 81). 
L. M. is supported by LabEX ENS-ICFP: ANR-10-LABX-0010/ANR-10-IDEX-0001-02 PSL*.
This research was supported in part by the National Science Foundation under Grant No. NSF PHY11-25915.


\appendix

\section{Spectral properties of the Lindbladian super-operator}
\label{app:Spectral}

In order to discuss the long-time properties of the dissipative dynamics, it is convenient to start 
from the spectral decomposition of the Lindbladian. Since $\mathcal L$ is in general 
a non-Hermitian operator, its eigenvalues are related to its Jordan canonical form~\cite{Kato}. 
Let us briefly review these results.
The Hilbert space of linear operators on the fermionic Fock space, $\mathbb H$, can be decomposed 
into the direct sum of linear spaces $\mathbb M_j$ (usually not orthogonal) such that if we denote 
with $\mathcal P_j$ the projectors onto such subspaces (usually not orthogonal)
and with $\mathcal N_j$ a nilpotent super-operator acting on $\mathbb M_j$, the following is true:
\begin{equation}
  \mathcal{L} = \sum_j \left[ \lambda_j \mathcal{P}_j + \mathcal N_j \right] .
\end{equation}
The $\{ \lambda_j \}$ are the generalized complex eigenvalues of the super-operator ${\mathcal L}$
and, for the case of a Lindbladian, have non-positive real part; 
the $\mathcal N_j$ can also be equal to zero. 
By this explicit construction it is possible to observe that the
$\{ \mathcal P_j\}$ and $\{ \mathcal N_j\}$ are all mutually commuting 
($\mathcal P_j \mathcal P_k = \delta_{j,k} \mathcal P_j$, $\mathcal P_j \mathcal N_k = \mathcal N_k \mathcal P_j = \delta_{j,k} \mathcal N_j$ and $\mathcal N_j \mathcal N_k = \delta_{j,k} \mathcal N_j^2$). 

Using these properties, the time evolution can be written as:
\begin{equation}
  \hat \rho(t) = e^{t \mathcal{L'}} [ \hat \rho(0) ]
  = \sum_j e^{\lambda_j t} e^{t \mathcal N_j} \mathcal{P}_j[\hat \rho(0)],
\end{equation}
which highlights that at a given time $t$ only the terms of the sum such that $|\Re(\lambda_j) \, t| \ll 1$ 
play a role. 
In the long-time limit, it is possible to represent the expectation value of any observable $\hat A$ as:
\begin{equation}
  \langle \hat{A}\rangle(t) \approx \text{tr} [ \hat{A} \, \mathcal P_0[\hat \rho(0)]] +\,
  e^{-\lambda_{\rm ADR} t} \, \text{tr} [ \hat{A} \,e^{t \mathcal N_{ \rm ADR}} \mathcal P_{ \rm ADR}[\hat \rho(0)]].
  \label{eq:steady:observable:app}
\end{equation}
Eq.~\eqref{eq:steady:observable:app} is the mathematical formula motivating 
Eq.~\eqref{eq:steady:observable} in the text, defining also the meaning of $\kappa$.

Let us mention that in the example discussed in the text $\mathcal N_{\rm ADR}=0$: 
this is observed by explicit inspection via exact diagonalization of small systems ($L=5$). 
Since the presence of a non-zero nilpotent super-operator is a fine-tuned property, 
it is reasonable to assume that the situation remains similar for longer systems.

\section{Analogies with the parent Hamiltonian}
\label{app:ParentHamiltonian}

\begin{figure}[t]
  \centering
  \includegraphics[scale=0.4]{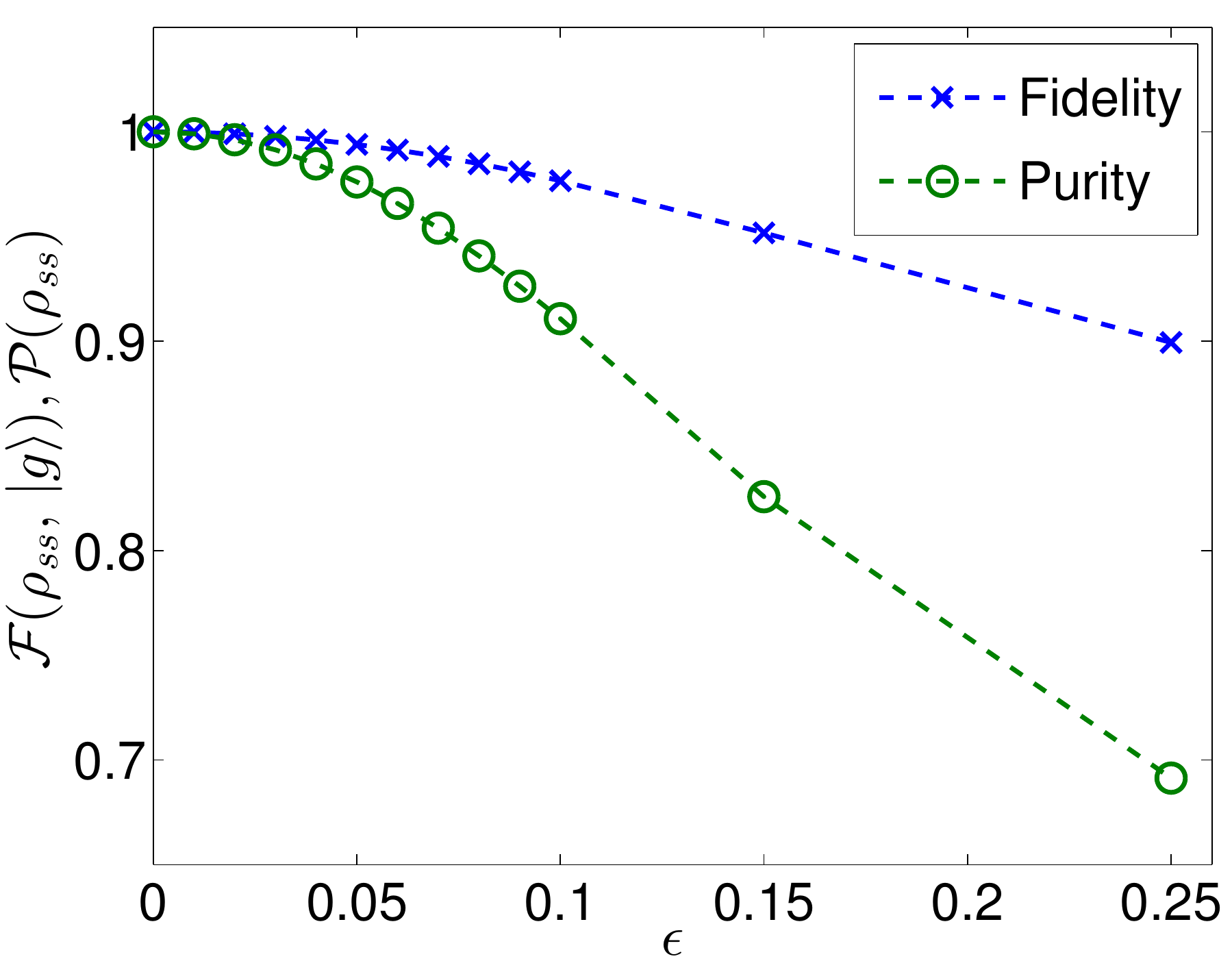}
  \caption{ (Color online) Fidelity $\mathcal{F}(\rho_{ss},\ket{g})$ and purity $\mathcal{P}(\rho_{ss})$
    for different values of $\epsilon$ in $\hat{L}'_{j, \epsilon}$ [see Eq.~\eqref{eq:pert:exc}].
}
  \label{appendix.fidelity.purity}
\end{figure}

\begin{figure}[t]
  \centering
  \includegraphics[scale=0.4]{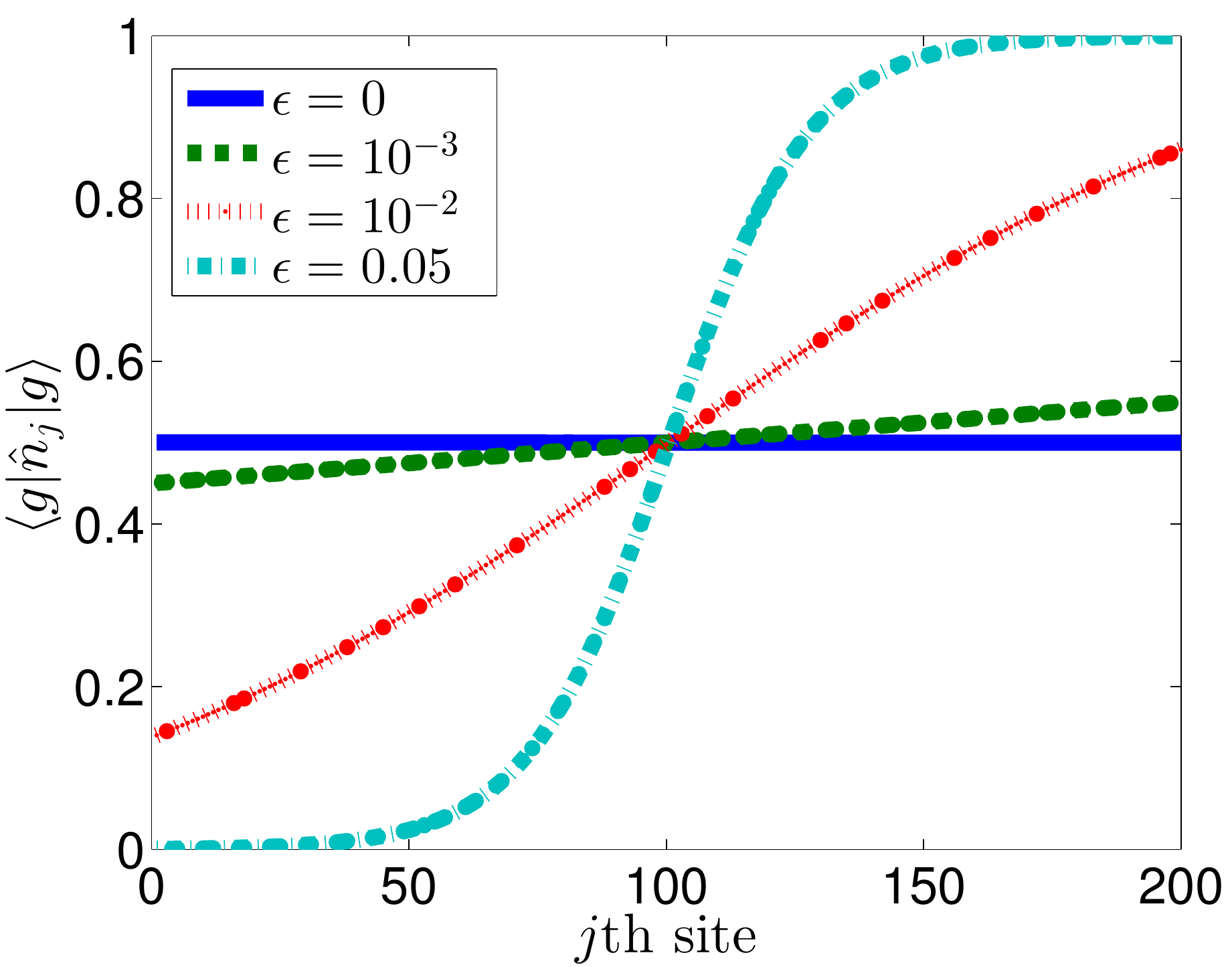}
  \includegraphics[scale=0.4]{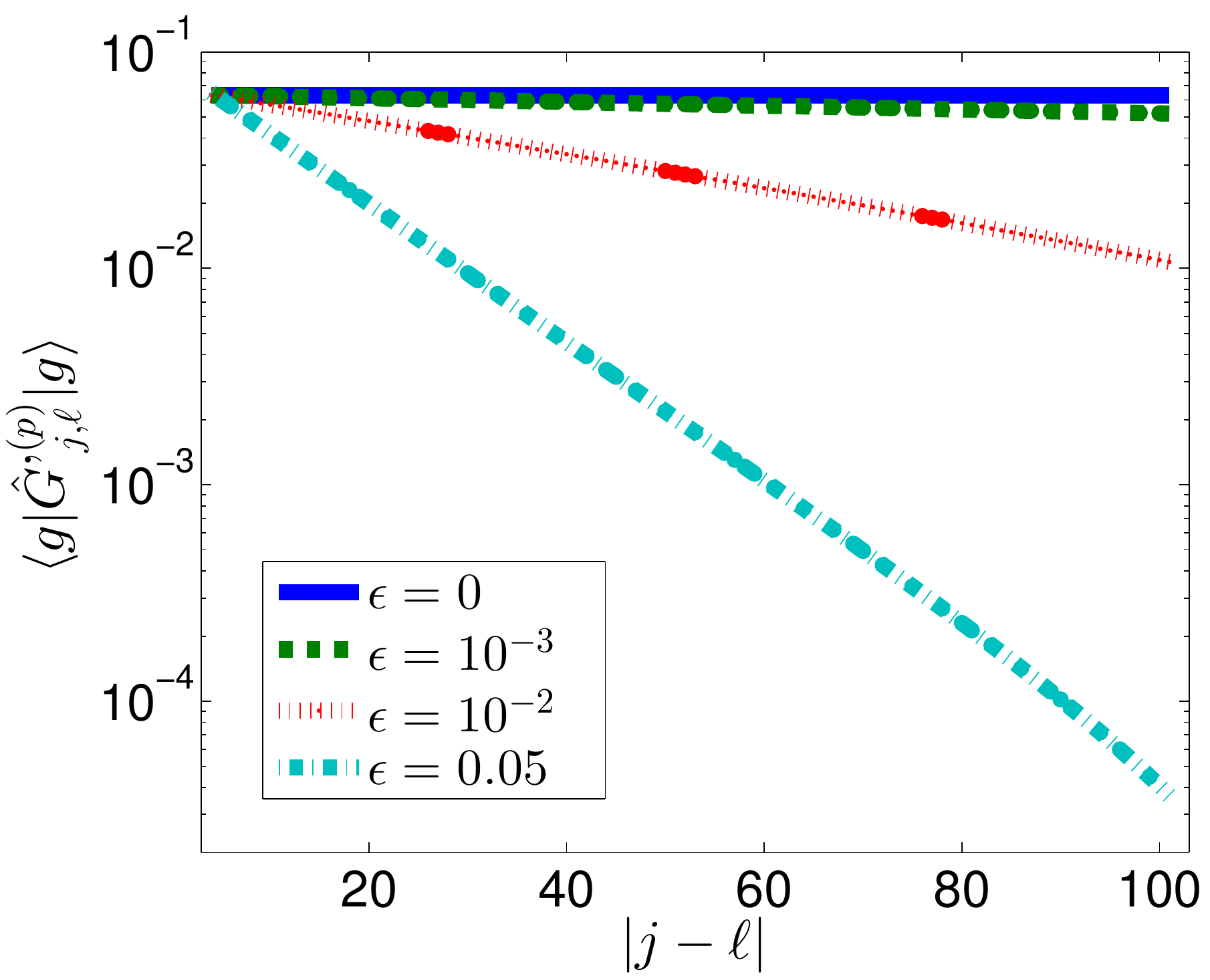}
  \caption{(Color online) (top) Density profile $\bra{g} \hat{n}_j \ket{g}$ 
    and (bottom) renormalized pairing correlations 
   $ \langle g| \hat O'^{(p) \dagger}_j \hat O'^{(p)}_{\ell} |g\rangle$, with $j=(L/2)-2 $ and $\ell>j$. 
    The computation is performed for a lattice with $L=200$ sites at half-filling and different values of $\epsilon$ in $\hat{L}'_{j, \epsilon}$ [see Eq.~\eqref{eq:pert:exc}].
}
  \label{appendix.mps.L40}
\end{figure}

In this Appendix we discuss some interesting analogies between the steady state $\hat \rho_{\rm ss}$ 
of the dissipative dynamics for the perturbed Lindblad operator $\hat L'_{j, \epsilon}$ 
in Eq.~\eqref{eq:pert:exc} with the ground state $\ket{g}$ of its parent Hamiltonian 
$\hat {\mathcal H}'_{p, \epsilon} = J \sum_j \hat L_{j, \epsilon}'^\dagger \hat L'_{j, \epsilon}$. 
It should be stressed that, since $\hat {\mathcal H}'_{p, \epsilon}$ does not have a zero-energy ground state,
there is no exact correspondence between both states.

We first study a small lattice with $L=8$ sites at half-filling, performing a Runge-Kutta integration
of the master equation. The initial state of the evolution is the ground state of $\hat \Ham_0$. 
In Fig.~\ref{appendix.fidelity.purity} it is shown that both the purity of the steady state 
$\mathcal{P}(\rho_{ss}) = {\rm tr} \big[\hat \rho_{ss}^2 \big]$ and 
its fidelity with the ground state of the parent Hamiltonian decrease with the perturbation strength. 
Notice, however, that for small perturbations the fidelity 
$\mathcal{F}(\hat \rho_{\rm ss}, \ket{g}) = \bra{g} \hat \rho_{\rm ss} \ket{g}$
remains close to one, thus revealing the similarity of the states in such regime. 

Such feature is also observed for larger lattices.
Using the MPDO method for $\hat \rho_{\rm ss}$ and an algorithm based on matrix product states
for $\ket{g}$, we analyze a lattice with $L=22$ sites at half-filling.
We compare the pairing correlations and density profiles for both states, which differ only 
for $\mathcal{O}(10^{-2})$, when the perturbation strength is $\epsilon \lesssim 0.05$ (not shown).
Let us explicitly show the results for the Hamiltonian case.
In Fig.~\ref{appendix.mps.L40} we show that, for a lattice with $L=40$ sites at half-filling, 
even a small perturbation ($\epsilon \sim 10^{-3}$) produces a non-negligible inhomogeneity. Moreover, the pairing correlations decay, indicating that such perturbation breaks the p-wave ordered nature of the purely dissipative dark state.

This similarity encourages the possibility of accessing some steady-state
properties for large lattices through the study of the ground states of the corresponding parent Hamiltonians, even if no mathematical connection is present and the mixedness of the state is expected to act like a finite temperature, washing out several ground-state properties.

\end{document}